\documentclass[preprint, 3p]{elsarticle}

\usepackage{preamble}

\begin{document}

\journal{Reliability Engineering and System Safety}

\begin{frontmatter}

\title{Niching Subset Simulation}

\author[ucl]{Hugh J. Kinnear}
\author[ucl]{F.A. DiazDelaO}

\affiliation[ucl]{organization={Clinical Operational Research Unit, Department of Mathematics},
            addressline={ UCL}, 
            city={London},
            postcode={WC1H 0BT}, 
            country={United Kingdom}}

\begin{abstract}
Subset Simulation is a Markov chain Monte Carlo method used to compute small failure probabilities in structural reliability problems. This is done by iteratively sampling from nested subsets in the input space of a performance function, i.e. a function describing the behaviour of a physical system. When the performance function has features such as multimodality or rapidly changing output, it is not uncommon for Subset Simulation to suffer from ergodicity problems. To address these problems, this paper proposes a new framework that enhances Subset Simulation with niching, a concept from the field of evolutionary multimodal optimisation. Niching subset simulation dynamically partitions the input space using support vector machines, and recursively begins anew in each set of the partition. A new niching technique, which uses community detection methods and is specifically designed for high-dimensional problems, is also introduced. It is shown that Niching Subset Simulation is robust against ergodicty problems and can also offer additional insight into the topology of challenging reliability problems.
\end{abstract}


\end{frontmatter}

\section{Introduction}\label{sec:introduction}

Reliability analysis is concerned  with calculating the probability of failure of a physical system, i.e. the probability that the system's demand exceeds its capacity. The system's behaviour is modelled by a performance function acting on an input space, which in practical applications is often high-dimensional. The function itself is frequently nonlinear and computationally expensive to evaluate. These properties preclude the use analytical or numerical integration techniques for calculating the probability of failure. Hence, specialised reliability methods have been developed. Reliability methods can be broadly sorted into one of, or commonly a combination of, three categories: (i) first and second order reliability methods \cite{amExactInvariantFirst1974,keshtegarHybridSelfadaptiveConjugate2018,derkiureghianSecondOrderReliability1987,huangNewDirectSecondorder2018} approximate the performance function using its Taylor expansion; (ii) surrogate-based methods learn the performance function using strategically chosen performance evaluations \cite{bucherFastEfficientResponse1990, kaymazApplicationKrigingMethod2005,bourinetRareeventProbabilityEstimation2016,cardosoStructuralReliabilityAnalysis2008}; and (iii) Monte Carlo-based methods estimate the probability of failure using stochastic model simulations \cite{schuellerEfficientMonteCarlo2009}.

When a system is well-designed, the probability of failure is expected to be small. When that happens, Monte Carlo estimators have a high variance. This means that a very large number of performance function evaluations are required for an accurate estimation. Variance reduction techniques such as \ac{IS} \cite{engelundBenchmarkStudyImportance1993}, directional sampling \cite{grootemanAdaptiveDirectionalImportance2011}, and line sampling \cite{pradlwarterApplicationLineSampling2007}, seek to mitigate this problem. \Ac{SuS}, originally proposed in \cite{auEstimationSmallFailure2001}, is a variance reduction technique that uses \ac{MCMC} algorithms to sample from a sequence of approximating conditional distributions \cite{auEstimationSmallFailure2001}. It is ideally suited for problems with implicitly defined, high-dimensional performance functions, and has been successfully applied in those contexts \cite{shekharProbabilisticAnalysisPiping2022,zhuSeismicSafetyEvaluation2024,zhangStudyStaticDynamic2023}. Many variants and improvements to \ac{SuS} have been suggested. Some are concerned with improving the effectiveness of the \ac{MCMC} procedure \cite{papaioannouMCMCAlgorithmsSubset2015,wangHamiltonianMonteCarlo2019,santosoModifiedMetropolisHastings2011,liaoInvestigationMarkovChain2024a,shieldsSubsetSimulationProblems2021a}, and others have combined \ac{SuS} with other reliability methods, such as \ac{IS} \cite{songSubsetSimulationStructural2009} and surrogate models \cite{huangAssessingSmallFailure2016,zhangMethodCombinedMetamodel2024,bourinetAssessingSmallFailure2011,papadopoulosAcceleratedSubsetSimulation2012,thalerReliabilityAnalysisComplex2024}.  \ac{SuS} is also a versatile algorithm, and has been adapted for use in other realms such as  Bayesian inference \cite{straubBayesianUpdatingStructural2015,diazdelaoBayesianUpdatingModel2017,betzBayesianInferenceSubset2018}, optimisation \cite{liSubsetSimulationUnconstrained2011}, and calibration through history matching \cite{gongHistoryMatchingSubset2021}. However, it has been found that certain performance functions cause \ac{SuS} to suffer from ergodicty issues and produce unreliable estimates for the probability of failure \cite{breitungGeometryLimitState2019}. Such examples tend to exhibit challenging features such as rapidly changing performance values and multiple local optima.

There have been several different approaches to adapting \ac{SuS} to make it more robust against ergodicity problems. Access to a surrogate model reduces the computational complexity of an \ac{MCMC} algorithm, and so facilitates an increased number of Markov chains, which in turn decreases the chance that they will collectively become stuck in a restricted region of the input space. However, modelling performance functions with the aforementioned challenging features can be difficult, and specific techniques have been developed. For example, spectral embedding-based reliability methods sequentially partition the input space and fit a separate model of the performance function in each set \cite{wagnerRareEventEstimation2022}. A \ac{SuS} Kriging scheme, devised in \cite{cuiImplementationMachineLearning2019a}, starts with an initial global Kriging model and is later decomposed into multiple local Kriging models using K-means clustering. Another approach is to directly alter the \ac{MCMC} procedure. The Modified Replica Exchange-Based \ac{MCMC} algorithm introduces an additional explorer chain which enables large moves between disparate areas of the input space \cite{sharmaModifiedReplicaExchangebased2023}. Subset simulation with fitness-based seed selection utilises alternative seed selection strategies in order to promote greater diversity amongst the Markov chain population \cite{abdollahiSubsetSimulationMethod2021}. Finally, sequential space conversion replaces the performance function with a series of control variates which are able to provide global information regarding sudden gradient changes and multiple local optima \cite{rashkiSESCNewSubset2021}. 

This paper uses ideas taken from the field of \ac{EMO} to mitigate the ergodicty issues of \ac{SuS}. In \ac{EMO}, \acp{EA} are repurposed for use in \ac{MMO}. \acp{EA} are a class of optimisation metaheuristics, inspired by the biological process of evolution. They can be broadly characterised as algorithms that maintain a population of individuals which evolve through the selection and breeding of parents to produce offspring. Notable examples of \acp{EA} include Genetic Algorithms \cite{hollandAdaptationNaturalArtificial1992,wrightGeneticAlgorithmsReal1991},  Evolution Strategies \cite{rechenbergEvolutionsstrategieOptimierungTechnischer1973,schwefelEvolutionsstrategieUndNumerische1975} and Differential Evolution \cite{stornDifferentialEvolutionSimple1997}. \ac{MMO} is the problem of finding multiple local optima of an objective function. This is distinct from an optimisation problem with a multimodal objective function, since in that case, even though local optima may be explored, the goal is to converge to one global optimum. 

\acp{EA} are natural candidates for \ac{MMO}, since the population of evolving individuals have the ability to explore the neighbourhoods of many local optima. Ultimately however, \acp{EA} are designed to converge to one optimal solution, given enough time. In \ac{EMO}, niching methods are applied to \acp{EA} in order to maintain individuals in the neighbourhood (called a niche) of many different local optima. The name niching is also biologically inspired, where a niche is a role a specific species plays in an environment. Preselection, the idea that offspring should replace their parents in the population, was the first niching method to be suggested \cite{cavicchioAdaptiveSearchUsing1970}, and since then a wide variety of techniques have been proposed. Fitness sharing \cite{goldbergGeneticAlgorithmsSharing1987} is based on the notion that if two individuals are close according to the Euclidean metric, then they are likely to be in the same niche. Index based methods, like ring topologies \cite{xiaodongliNichingNichingParameters2010}, create a network through which individuals can communicate. Hill-valley detection
\cite{ursemMultinationalEvolutionaryAlgorithms1999} considers the topology of the objective function between individuals.

This paper makes two original contributions. The first is \ac{NSuS}, a general framework that combines \ac{SuS} with niching methods by using \acp{SVM} to dynamically partition the input space. The bridge between fields that \ac{NSuS} provides enables reliability analysis to benefit from the wealth of ideas in the \ac{EMO} literature. However, one of the main advantages of \ac{SuS} is its ability to perform well in high dimensions, whilst research on niching methods tends to focus on low dimensional objective functions \cite{liBenchmarkFunctionsCEC2013}. This leads to the second contribution, the \ac{HVG}, a new niching method that uses \ac{ALP} \cite{raghavanLinearTimeAlgorithm2007}, and is specifically designed to work well in high dimensions. The primary motivation for \ac{NSuS} is to prevent ergodicity problems, but it also offers other benefits. \ac{SuS} produces an estimate for the probability of failure, but offers little insight into the underlying reliability model. In contrast, \ac{NSuS} is able to return the location of multiple failure modes and quantify their individual contributions to the probability of failure. This additional information can be practically useful for many reasons. For an engineer developing a design to improve the safety of a system it is useful to know the dominant failure modes. If a practitioner has many similar reliability problems, understanding the topology of one may simplify the analysis of the rest.

The paper is organised as follows. Section \ref{sec:subset simulation} formally defines the probability of failure and gives an overview of the original \ac{SuS} algorithm. \ac{NSuS} and its theoretical framework are covered in Section \ref{sec:niching subset simulation}. The heuristics governing \ac{NSuS} and the \ac{HVG} are introduced in Section \ref{sec:the partitioner}. In Section \ref{sec:numerical examples}, illustrative numerical examples are presented, alongside more practical oscillator reliability problems. The paper is summarised and concluded in Section \ref{sec:conclusion}, including ideas for future directions of research.

\ifSubfilesClassLoaded{%
    \newpage 
    \bibliography{references}%
}

\end{document}

\section{Subset Simulation}\label{sec:subset simulation}

\subsection{The Original Algorithm}\label{sec:the original algorithm}

In reliability analysis, the performance function $g\colon \R^{d} \rightarrow \R$ assigns a scalar performance to every combination of input values in the input space $\R^{d}$. The failure region is defined by a critical threshold $b \in \R$, such that $F = \{\bm{x} \in \R^{d} : g(\bm{x}) \geq b\}$. The input space is endowed with an input distribution, determined by a probability density function $f(\cdot)$, such that the probability of failure is defined as:
\begin{equation}\label{eq:failure proability}
    P_{F} = \int_{\R^{d}} \one_{F}(\bm{x})f(\bm{x})d\bm{x},
\end{equation}
where $\one_{F}(\cdot)$ is an indicator function. Without loss of generality, in this paper, the critical threshold is assumed to be $0$. To estimate the probability of failure, the original \ac{SuS} algorithm models the failure region using a nested sequence of intermediate failure regions. The idea is to sequentially approximate a rare event using relatively frequent events which are easier to sample from. Each intermediate failure region has an associated level contained within in it, where a level is a member of the set  $\mathcal{L} = \{\bm{X} \subset \R^d : |\bm{X}| = n \}$ and $n \in \N$ is the level size.

The initial level, $\bm{X}^{0}$, is created by sampling $n$ times from the input distribution. The performance function $g(\cdot)$ is then evaluated at each sample in the the initial level. The top $n_{c} = np$ performing samples are chosen as seeds for the next level, where the level probability $p\in (0,1]$ is a parameter chosen by the modeller. Note that this requires $n$ and $p$ to be chosen such that $n_c$ is an integer.  A fixed level probability of $p=0.1$ is used in all the examples of this paper, since it has been shown in the literature \cite{zuevBayesianPostprocessorOther2012} that $p \in [0.1,0.3]$ is optimal. The performance of the lowest performing seed becomes the first intermediate threshold $b_1$, which defines the first intermediate failure region $F_{1} = \{\bm{x} \in \R^{d} : g(\bm{x}) \geq b_{1}\}$. A Markov chain of length $n_s = p^{-1}$, with stationary distribution $f(\bm{x})\one_{F_1}(\bm{x})$, is then created from each of the seeds. Again, note that $p$ must be chosen so that $n_s$ is an integer. The Markov chains taken together comprise the next level, $\bm{X}^{1}$. The process is repeated to create all subsequent levels. Namely, to create $\bm{X}^{k}$, seeds are chosen from $\bm{X}^{k-1}$. The seeds define an intermediate threshold $b_{k}$ and an intermediate failure region $F_{k}$. The level is then created from the seeds using Markov chains with a stationary distribution of $f(\bm{x})\one_{F_k}(\bm{x})$. This process continues until a stopping condition is met. The output of the algorithm is the sequence of levels it creates, $(\bm{X}^k)_{k=0}^{m}$.

Since the standard Metropolis \ac{MCMC} algorithm struggles in high dimensions \cite{katafygiotisGeometricInsightChallenges2008}, many alternative \ac{MCMC} algorithms for use with \ac{SuS} have been developed through the years \cite{papaioannouMCMCAlgorithmsSubset2015}. This paper uses the modified Metropolis algorithm \cite{auEstimationSmallFailure2001} (summarised in Algorithm \ref{alg:modified metropolis}). The modified Metropolis algorithm takes advantage of the special structure of the probability density functions which \ac{SuS} targets, and so it is able to have good sampling efficiency in high dimensions. To reflect the different possible choices for a Markov chain algorithm, a generic notation will be used. Define a function $\textsc{Step}_f: \R^{d} \rightarrow \R^{d}$, where $f$ is is the stationary distribution of the Markov chain. That way, given a starting sample $\bm{x} \in \R^{d}$, the next member of the chain will be $\textsc{Step}_f(\bm{x})$. The structure of the stationary distribution in Algorithm \ref{alg:modified metropolis} assumes independence between the inputs. This is due to the convention of assuming the input distribution is a standard multivariate normal. This convention is justified by altering the performance function appropriately depending on the available description of the input distribution. For instance, if the input distribution is explicitly known, then the Rosenblatt transformation \cite{rosenblattRemarksMultivariateTransformation1952} can be applied. Alternatively, if only the marginal distributions and correlations are given the joint probability distribution can be approximated by a Nataf distribution \cite{natafDeterminationDistributionsDont1962}. 
\begin{algorithm}
\caption{Modified Metropolis}\label{alg:modified metropolis}
\textbf{Input} \\
Sample: $\bm{x} \in \R^d $ \\
Target distribution: $f(\bm{x}) = \one_{F}(\bm{x}) \prod f_{i}(x_i)$ \\
Proposal distribution: $q(\cdot|\bm{x})$
\begin{algorithmic}[1]
\Procedure{ModifiedMetropolis}{$\bm{x}$}
    \For{$1\leq i \leq d$}
        \State Sample $ x'_{i} \sim q(\cdot|x_i)$
        \State $\theta \gets \min(1, f(x'_i)/f(x_i))$
        \State $\alpha \gets \textsc{bernoulli}(\theta)$
        \If{$\alpha = 0$}
            \State $x'_{i} \gets x_i $
        \EndIf
    \EndFor
    \State $\bm{x}' \gets (x'_1,\dots,x'_d)$
    \If{$\one_{F}(\bm{x}) = 0$}
        \State $\bm{x}' \gets \bm{x} $ 
    \EndIf
    \State \textbf{return} $\bm{x}'$
\EndProcedure
\end{algorithmic}
\end{algorithm}

The \ac{SuS} algorithm stops when enough samples have been produced in the failure region. Explicitly, the stopping condition is activated on level $m$ if
\begin{equation}\label{eq:fail stop}
    \sum_{\bm{x} \in \bm{X}^{m}} \one_{F}(\bm{x}) \geq n_c.
\end{equation}
There are degenerate cases where \ac{SuS} fails to ever produce failure samples, and so an additional stopping condition is required. The examples in this paper check if the next proposed intermediate threshold is the same as the previous intermediate threshold.  Again, to reflect the potential for different possible stopping conditions, a generic notation will be used: $\textsc{Stop}: \mathcal{L} \rightarrow \{\textbf{True},\textbf{ False}\}$. Note that a stopping condition may not strictly only act on a particular level, for example it may consider the total numer of performance function evaluations. The complete \ac{SuS} procedure is summarised in Algorithm \ref{alg:subset simulation}.

\begin{algorithm}
\caption{Subset Simulation}\label{alg:subset simulation}
\textbf{Input} \\
Level size: $n \in \N $ \\
Level probability: $p \in (0,1]$ \\
Input distribution: $f\colon\R^{d} \rightarrow \R$ \\ 
Performance function: $g\colon\R^{d} \rightarrow \R$ \\
\textbf{Definitions} \\ 
$n_c \gets np$ \\
$n_s \gets p^{-1}$ \\
\textbf{Subroutines} \\
\textsc{Stop}:  $\mathcal{L} \rightarrow \{\textbf{True},\textbf{ False}\}$ \\ 
\textsc{Step}: $\R^d \rightarrow \R^d$
\begin{algorithmic}[1]
\Procedure{SubsetSimulation}{}
    \State Sample $\bm{x}^0_1,\dots,\bm{x}^0_n \sim f$
    \State $\bm{X}^0 \gets (\bm{x}^0_i)_{i=1}^n$
    \State $m \gets 0$
    \While{$\textsc{Stop}(\bm{X}^m)$ is \textbf{False}}
        \State let $\bm{x}'_1,\dots,\bm{x}'_n$ be a relabelling of $\bm{X}^m$ such that $g(\bm{x}'_1) \geq \dots \geq g(\bm{x}'_n)$
        \State $m \gets m +1$
        \State $b_m \gets g(\bm{x}'_{n_c}) $
        \State $ F_{m} \gets \{\bm{x} \in \R^{d}: g(\bm{x}) \geq b_m\}$
        \State $f_m(\bm{x}) \gets f(\bm{x})\one_{F_m}(\bm{x})$
        \For{$ 1 \leq i \leq n_c $}
            \State $\bm{x}^{m}_{i,1} \gets \bm{x}'_{i}$
            \For{$ 2 \leq j \leq n_s $}
                \State $\bm{x}^{m}_{i,j} \gets     \textsc{Step}_{f_m}(\bm{x}^{m}_{i,j-1})$
            \EndFor
        \EndFor
        \State $\bm{X}^m \gets ((\bm{x}^m_{i,j})_{i=1}^{n_c})_{j=1}^{n_s}$
    \EndWhile
    \State \textbf{return} $(\bm{X}^k)_{k=0}^{m}$
\EndProcedure
\end{algorithmic}
\end{algorithm}

\subsection{Estimating the Probability of Failure}\label{sec:estimating probability of failure}

The output levels of \ac{SuS} can be used to estimate any failure probability $P_{F}$ defined by a threshold $b$ and failure region $F$. In particular, $b$ need not be the critical threshold. Let $b_0 = -\infty$ and  $m'=\max (\{k :b_{k} < b  \})$. Since $F \subseteq F_{m'} \subseteq  \dots \subseteq F_{1} \subseteq F_{0} = \R^d$, the product rule can be used to obtain an expression for the failure probability:
\begin{equation}\label{eq: failure probability decomp}
    P_{F} = \prob(F_1|F_0)\prob(F_2|F_1)\dots \prob(F_{m'}|F_{m'-1})\prob(F|F_{m'}).
\end{equation}
It is therefore natural to estimate $P_{F}$ as the product of estimators of each term in Equation \ref{eq: failure probability decomp}. These estimators are denoted here as $\hat{P}_1, \hat{P}_2,\dots,\hat{P}_{m'+1}$. The following definition is kept general for convenience, but the case described above can be recovered by letting $m=m'$ and $F_{m+1} = F$.

\begin{definition}[SuS Estimator]\label{def:sus estimator}
Given a nested sequence of sets $F_0,F_1,\dots,F_{m},F_{m+1}$ and a sequence of levels $\bm{X}^0, \dots, \bm{X}^m$ each of size $n$, an estimator $\hat{P}$ is a \ac{SuS} estimator for $\prob(F_{m+1}|F_0)$ with respect to the density function $f$ if it has the following product form:
\begin{equation}\label{eq:generic product}
    \hat{P} = \prod_{i=1}^{m+1} \hat{P}_{i},
\end{equation}
where $\hat{P}_{i} = \displaystyle\frac{1}{n} \sum_{\bm{x} \in \bm{X}^{i-1}} \one_{F_{i}}(\bm{x})$ and $\bm{x} \sim f|F_i$ for all $\bm{X}^i$.
\end{definition}

If it is assumed that the intermediate thresholds are chosen a priori rather than dynamically, and that the samples generated by different chains are uncorrelated through the indicator functions, it can be shown that \ac{SuS} estimators are consistent and asymptotically unbiased \cite{auEstimationSmallFailure2001}. Based on the sequence of levels $\bm{X}^0, \dots, \bm{X}^{m'}$ a \ac{SuS} estimator can be used to estimate the required failure probability:
\begin{equation}\label{eq:sus_estimator}
    P_F = \prob(F) = \prob(F|F_0) \approx \hat{P}_{F} =  \prod_{i=1}^{m'+1} \hat{P}_{i} = p^{m'} \hat{P}_{m'+1},
\end{equation}
where the last equality holds since $\hat{P}_{k} = p $ for $1 \leq k \leq m'$ due to the adaptive choice of intermediate thresholds.

It is straightforward to see that the first estimator in the product, $\hat{P}_{1}$, is a \ac{DMC} estimator. The corresponding \ac{CoV} is given by:
\begin{equation}\label{eq:delta_1}
    \delta_1 = \sqrt{\frac{1-P_{1}}{nP_{1}}}.
\end{equation}
An estimation of the \ac{CoV}, $\hat{\delta}_1$, can be made by substituting $\hat{P}_{1}$ for $P_1$ in Equation \ref{eq:delta_1}.

The rest of the estimators in the product in Equation \ref{eq:sus_estimator} are \ac{MCMC} estimators. Normally, Markov chain methods require a burn-in period before the samples have the target distribution. \ac{SuS} however has a property known as perfect sampling, where the seeds of the chains will always have been sampled according to the target distribution. This means that there is no burn-in period and all the samples can be used in the calculation. The \ac{CoV} of the MCMC estimators is given by
\begin{equation}\label{eq:delta_k}
    \delta_k = \sqrt{\frac{1-P_k}{nP_k} (1+\gamma_k)},
\end{equation}
where $\gamma_k$ is a factor accounting for the correlation between samples in the same chain. An estimator for the \ac{CoV}, $\hat{\delta}_k$ for $2 \leq k \leq m' + 1$, can be obtained by substituting $\gamma_k$ for an estimation $\hat{\gamma_k}$ and $P_k$ for $\hat{P}_{k}$ in Equation \ref{eq:delta_k} \cite{auEstimationSmallFailure2001}.

Since the samples generated by Markov chains are correlated, \ac{MCMC} estimators will have a higher \ac{CoV}. than \ac{DMC} estimators given the same number of samples. The effective sample size of an \ac{MCMC} estimator is $n/(1+\gamma_k)$. The higher the correlation of the chains, the lower the effective number of samples. A common estimator used for the \ac{CoV} of the SuS estimator $\delta$ is
\begin{equation}\label{eq:sus cov}
    \hat{\delta} = \sqrt{\sum_{k=1}^{m+1} \hat{\delta}_i^2}.
\end{equation}
\ifSubfilesClassLoaded{%
    \newpage 
    \bibliography{references}%
}

\end{document}

\section{Niching Subset Simulation}\label{sec:niching subset simulation}

\subsection{Motivation}\label{sec:motivation}

A common object of study in the reliability literature is the design point. Let the limit state surface be the set $\Lambda = \{\bm{x} \in \R^{d} : g(\bm{x}) = b\}$. The design point is then defined as $x \in \Lambda$ closest to the origin when the input variables have been transformed to standard normal space. It is possible for a failure region to have multiple design points. Equivalently, a design point is a point in the failure region with the largest possible probability density, and so the neighbourhood of a design point makes a relatively large contribution to the probability of failure. Due to this, it is vital that \ac{SuS} is able to produce samples in the neighbourhood  of design points so that it can make a reliable estimate of the probability of failure. If \ac{SuS} fails to do so, it will likely underestimate the true probability of failure.

As discussed in the previous section, at any given level of a \ac{SuS} run, the highest performing samples are chosen as seeds. SuS then defines the next level by exploring the input space in the neighbourhood of those seeds. This greedy approach is often a sensible course of action. However, there are cases where unimportant sets in earlier intermediate failure regions become important in later intermediate failure regions. Under these circumstances, it is possible that \ac{SuS} produces no seeds in the now important set. The two-dimensional piecewise linear function described in \cite{breitungGeometryLimitState2019} was constructed to highlight exactly this type of deficiency of the \ac{SuS} algorithm. This performance function is given by
\begin{equation}\label{eq:piecewise linear}
\begin{aligned}
    & g(\bm{x}) = - \min(g_1(x_1),g_2(x_2)), \,\text{where}\\
    & g_1(x_1) = 
    \begin{cases}
        4 - x_1 & x_1 > 3.5, \\
        0.85 - 0.1x_1 & x_1 \leq 3.5,
    \end{cases} \\
    & g_2(x_2) = 
    \begin{cases}
        0.5 - 0.1x_2 & x_2 > 2, \\
        2.3 - x_2 & x_2 \leq 2.
    \end{cases} 
\end{aligned}
\end{equation}
For consistency, note that Equation \ref{eq:piecewise linear} has been multiplied by $-1$, since this paper uses the convention that \ac{SuS} attains progressively higher intermediate thresholds.

\begin{figure}
    \centering
    \begin{subfigure}[b]{\textwidth}
        \includegraphics[scale=0.7]{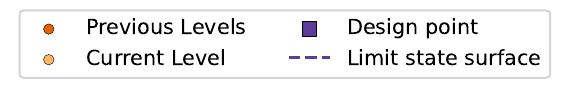}
    \end{subfigure}
    \begin{subfigure}[b]{0.475\textwidth}
        \centering
        \includegraphics[scale=0.55]{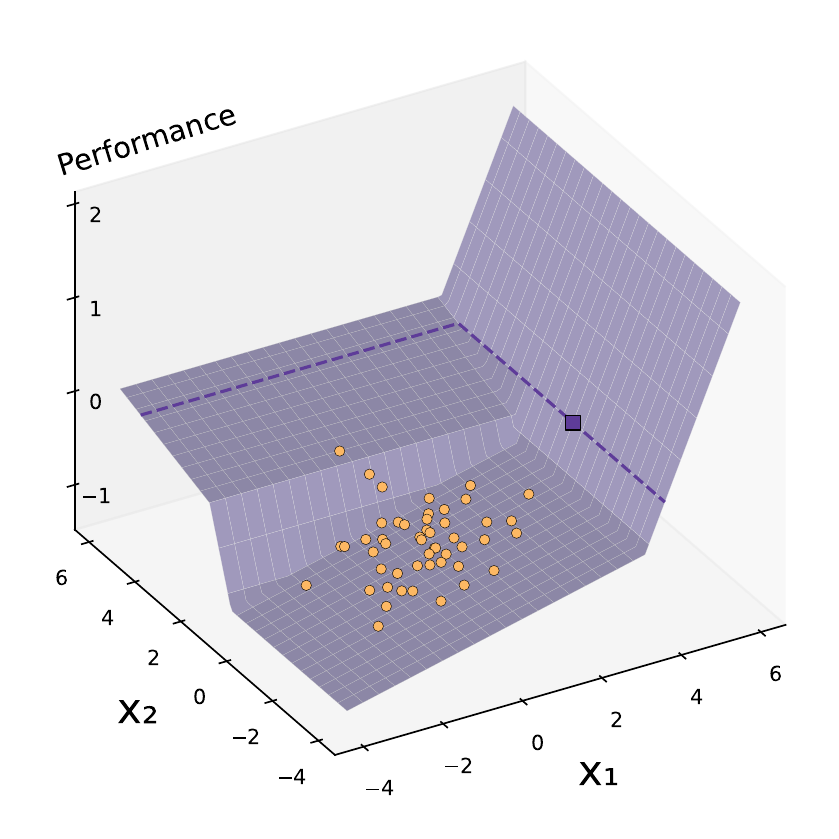}
        \caption{Initial level}
        \label{fig:pwl_3d_sus_1}
    \end{subfigure}
    \begin{subfigure}[b]{0.475\textwidth}
    \centering
        \includegraphics[scale=0.55]{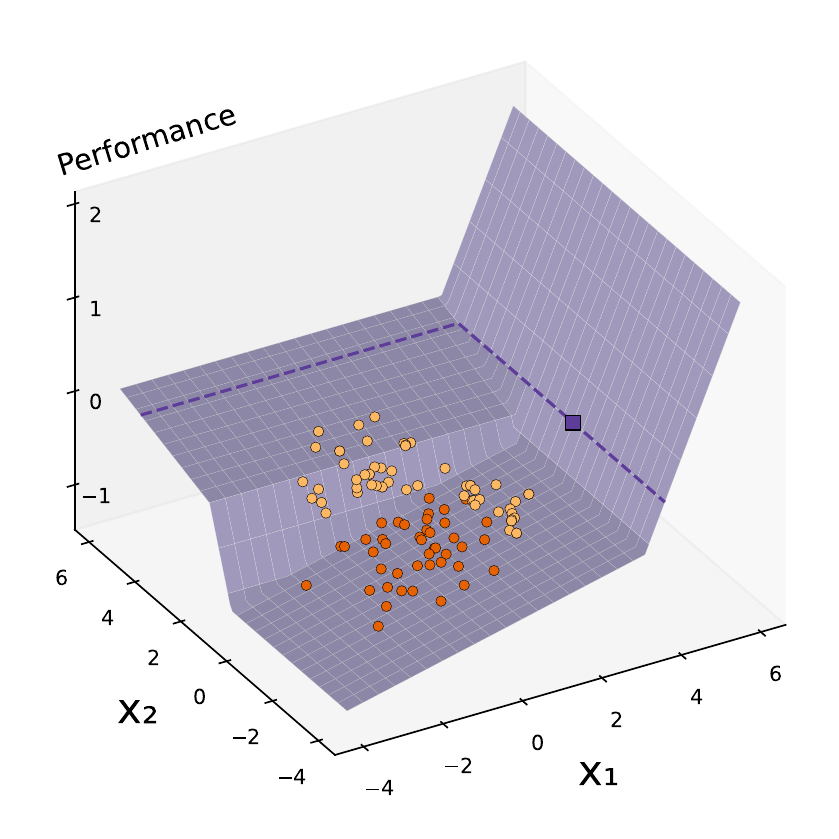}
        \caption{Level $1$}
        \label{fig:pwl_3d_sus_2}
    \end{subfigure}
    \begin{subfigure}[b]{0.475\textwidth}
        \centering
        \includegraphics[scale=0.55]{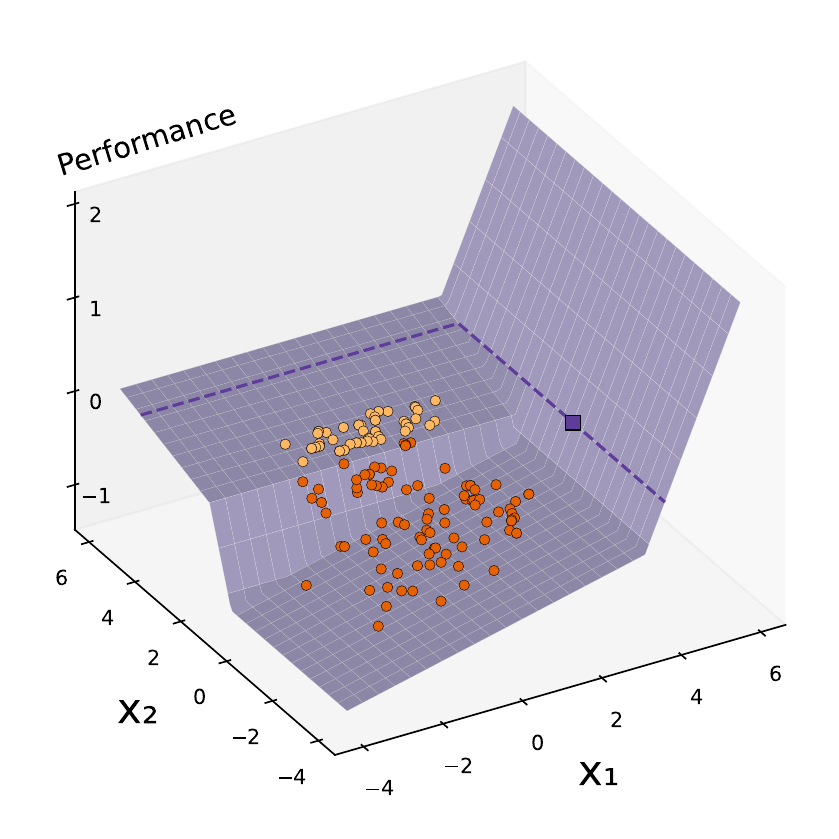}
        \caption{Level $2$}
        \label{fig:pwl_3d_sus_3}
    \end{subfigure}
    \begin{subfigure}[b]{0.475\textwidth}
        \centering
        \includegraphics[scale=0.55]{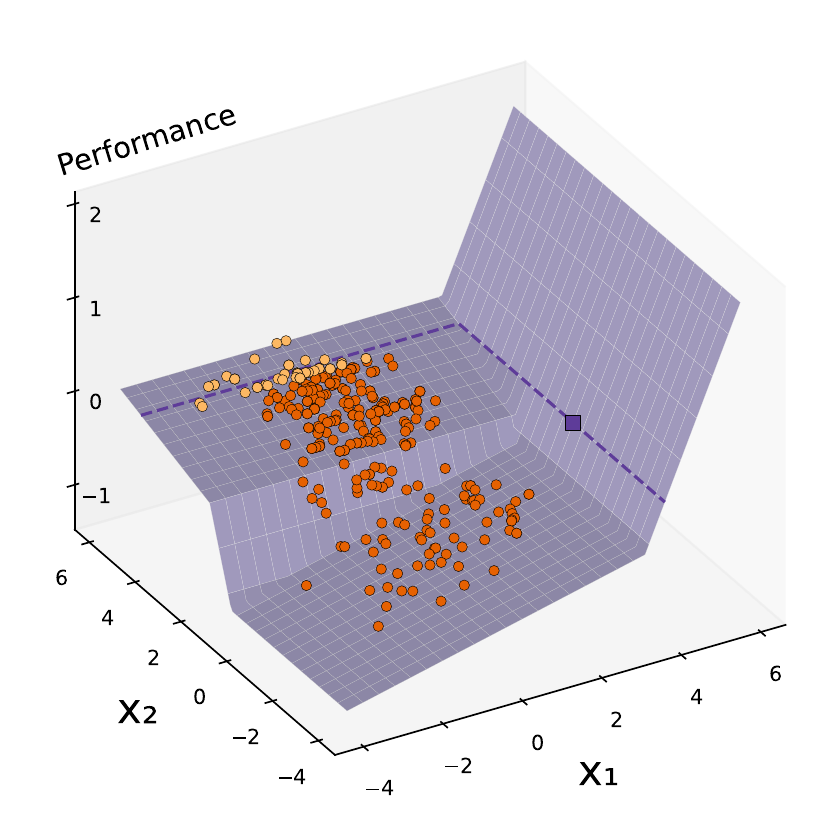}
        \caption{Level $6$}
        \label{fig:pwl_3d_sus_7}
    \end{subfigure}
    \caption{SuS running on the piecewise linear function. For clarity, only every $10^{th}$ sample has been plotted. SuS is lead away from the design point resulting in a poor estimate for the probability of failure.}
    \label{fig:sus piecewise}
\end{figure}

A single run of \ac{SuS} acting on the piecewise linear function is shown in Figure \ref{fig:sus piecewise}. In the region populated by the initial level, the highest performing samples are in the direction of positive $x_2$, since it has steepest gradient, and so this is the direction in which \ac{SuS} travels. However, further from the mean of the input distribution, the performance function begins to increase more rapidly in the direction of positive $x_1$, meaning that ultimately this is where the design point is located. Eventually, SuS manages to produce some samples in the failure region and the stopping condition is satisfied. Despite this, the highest density area of the failure region has not been explored, and the probability of failure will be severely underestimated. Since SuS is a stochastic algorithm, there will be runs where the neighbourhood of the design point will be sampled from. In those cases, however, the \ac{SuS} estimator will tend to overestimate the probability of failure.

\begin{figure}
    \centering
    \includegraphics[scale=0.5]{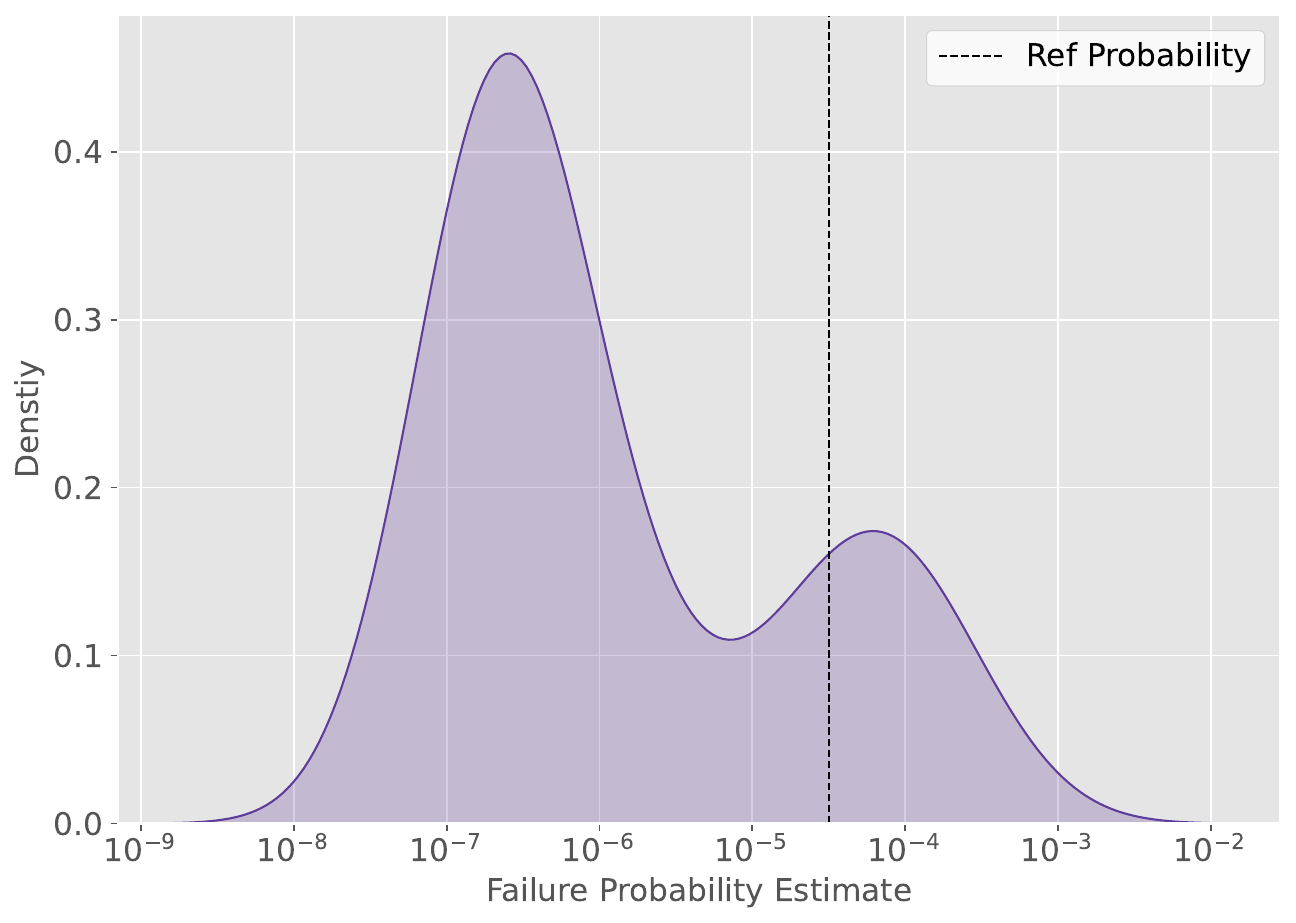}
    \caption{Kernel density estimate for 100 SuS probability of failure estimators on the piecewise linear function, compared to a reference probability.}
    \label{fig:linear_sus_kde}
\end{figure}

To see how this behaviour leads to an undesirable statistical estimator, Figure \ref{fig:linear_sus_kde} shows a kernel density estimate of the SuS estimator for the piecewise linear function built from 100 \ac{SuS} runs with level size $n=1000$. A logarithmic scale has been used since the range of estimates of the probability of failure span multiple orders of magnitude. For this example, a reference probability of $3.18 \times 10^{-5}$ was calculated using \ac{DMC} with $10^8$ samples. The figure shows a bimodal probability density function. This is unsurprising: the right mode corresponds to the 65 degenerate runs that do not populate the neighbourhood of the design point, whereas the left mode is the result of 35 runs that do. The empirical mean of all the failure probability estimators is  $2.95 \times 10^{-5}$. Whilst this could be considered to be a decent estimate of the reference probability, the variance of the \ac{SuS} estimator is prohibitively large for practical purposes. Additionally, there is very little density in the neighbourhood of the reference reliability itself, and so it is very unlikely that any individual \ac{SuS} run will serve as a useful estimate for the reliability. 

The simplest approach to overcoming these issues is to adjust the parameters of \ac{SuS}. The number of seeds can be increased by either increasing the level size or the level probability. This will help, since more seeds means more of the input space will be explored. Another option is to increase the spread of the proposal distribution which would allow the chains to move between disparate regions of the input space more freely. Both of these approaches share a fundamental flaw. In the case where the performance function is a black box, it is impossible to know a priori how these parameters should be adjusted. If the level size or level probability is too large, needless computational cost will be incurred. If the spread of the proposal distribution is too large, the correlation of the chains may increase, which in turn causes the \ac{CoV} of the probability of failure estimate to increase. To address all of these difficulties, this paper proposes \ac{NSuS}.

\subsection{Proposed Algorithm}\label{sec:proposed algorithm}

\ac{NSuS} is a general framework, based on the original SuS algorithm, that aims at improving the exploration of the input space. The idea is to create a partition of the input space at each level. Then, in each set of the partition, a new level is created using the highest performing samples in that set as seeds. The process continues recursively on all the new levels. Whilst \ac{SuS} creates levels in series and chooses seeds as the globally highest performing samples, \ac{NSuS} creates parallel levels and chooses the locally best performers as seeds.

The partitions of the input space are created by a function called the partitioner. Of course, the choice of partitioner is critical and the details will be provided in Section \ref{sec:the partitioner}. For now, a generic partitioner is defined as: $\textsc{Partition}: \mathcal{L} \rightarrow \textbf{Part}(\R^{d})$, where $\textbf{Part}(\R^{d})$ denotes the set of all possible partitions of $\R^d$. During an \ac{NSuS} run, the partitioner is used on each level after it has been created. If a partition consisting of one set is returned, where that set is necessarily the entire input space, the algorithm carries on as traditional \ac{SuS} would. If a proper partition of the input space is returned, say $A_1,\dots,A_{n_p}$, then the algorithm branches. There will be an independent \ac{NSuS} run, called a branch, started in each set of the partition.

Let $\bm{X}$ be the level that caused the branching. For each $A_i$ with $1 \leq i \leq n_p$, the algorithm treats $A_{i} \cap \bm{X}$ as the initial level of the new branch, and creates the next level of the branch in a nearly identical manner to SuS. The only difference is that the amount of computational budget allocated to each branch now must be decided. In practice this means the number of seeds, and chain length, must be decided for each branch. The approach taken in this paper is to evenly split amongst all the branches those computational resources that would be assigned to one level in the original  \ac{SuS} algorithm. That is, $n(1-p)$ attempted Markov chain steps. The number of seeds chosen in each branch is equal, where possible, to the number of seeds that would have been chosen if there were no split. The chain length is then adjusted in each branch to share out the computational budget. Formally,
\begin{equation}\label{eq:allocate}
    n_c = \min(\floor*{|\bm{X}|p},|A_{i} \cap \bm{X}|), \qquad n_s = \floor*{\frac{n(1-p)}{N_p n_c}} + 1,
\end{equation}
where $\floor{\cdot}$ is the floor function and $N_p$ is the total number of branches. For a partition of one set, and one total branch, this allocation simplifies to the allocation made by \ac{SuS}. Of course there are many possible alternatives for assigning computational resources.

The Markov chains in each branch are confined to their respective sets using the indicator function $\one_{A_i} (\bm{x}) $. Note that the indicator function also applies to any branches that this branch might create. From then on, independent \ac{NSuS} runs take place in each branch. The output of the algorithm is a tree structure $T$ that belongs to the space of all possible trees $\mathcal{T}$. The nodes of $T$ are triples of the form $(\bm{X},A,F)$ where $\bm{X}$ is a level, $A \subset \R^d$ and $F$ is a failure region. The root of the tree is $(\bm{X}^0,\R^d,\R^d)$ where $\bm{X}^0$ is the initial level. If a level $\bm{X}$ is created using seeds taken from  $\bm{X}'$, then $\bm{X}$ is a child of $\bm{X}'$. Let a leaf be the last created level of a branch. Let the depth of a node be the number of nodes in the path from the itself to the root, excluding the root.

The \ac{NSuS} framework offers new options for stopping conditions. Each branch can be considered an independent SuS run and consequently any combination of stopping conditions designed for \ac{SuS} can be applied to any branch. The most straightforward strategy is to stop the entire algorithm once each individual branch has been stopped. However, there is now the additional possibility of stopping conditions that consider all the branches in parallel. For example, consider the case where some branches have reached the failure region and have triggered a stopping condition, whilst others are still searching for the failure region. It may become clear that the maximum possible contribution to the overall estimate of those branches still searching is too small to be relevant, and so they should be stopped. In this case, the stopping condition would have to monitor all branches simultaneously. Formally, the stopping conditions now act on $\mathcal{T}$ rather than $\mathcal{L}$.

In the original \ac{SuS} algorithm, the level created last is always chosen as the next level to update. In contrast, \ac{NSuS} must decide between the current branches as to which level will be updated. The numerical examples in this paper use the basic strategy of always choosing the branch with the lowest depth until a stopping condition is triggered. Another simple approach would be to choose the branches uniformly at random. More sophisticated strategies are certainly possible. For instance, the size of the failure probability in each branch could be estimated, and then the branch with largest section could be chosen first. Of course, if a particular \ac{NSuS} framework is set up in such a way that the branches do not impact one another, the order in which the branches are updated is irrelevant. To reflect that there are many possible choices, a generic choice function will be defined: $\textsc{Choose}:\mathcal{T} \rightarrow \mathcal{L}$. The \ac{NSuS} procedure is summarised in Algorithm \ref{alg:niching subset simulation}. There is one minor implementation detail. It is possible, though uncommon, for a set of a partition to contain no samples, or for a branch to be allocated a chain length of $1$. In these cases the branch should be stopped automatically.

\begin{algorithm}
\caption{Niching Subset Simulation}\label{alg:niching subset simulation}
\textbf{Input} \\
Minimum level size: $n \in \N $ \\
Level probability: $p \in (0,1]$ \\
Input distribution: $f\colon \R^{d} \rightarrow \R$ \\ 
Performance function: $g\colon \R^{d} \rightarrow \R$ \\
\textbf{Definitions} \\ 
$n_s \gets p^{-1}$ \\
\textbf{Subroutines} \\
\textsc{Stop}:  $\mathcal{T} \rightarrow \{\textbf{True},\textbf{ False}\}$ \\ 
\textsc{Step}: $\R^d \rightarrow \R^d$\\ 
\textsc{Partition}: $\mathcal{L} \rightarrow \textbf{Part}(\R^d)$ \\
\textsc{Choose}: $\mathcal{T} \rightarrow \mathcal{L} $
\begin{algorithmic}[1]
\Procedure{NichingSubsetSimulation}{}
    \State $N_p \gets 1$
    \State Sample $\bm{x}_1,\ldots,\bm{x}_n \sim f(x)$ 
    \State $\bm{X} \gets (\bm{x}_i)_{i=1}^n$
    \State Make $(\bm{X},\R^d,\R^d)$ root of $T$
    \While{$\textsc{Stop}(T)$ is \textbf{False}}
        \State $\bm{X},A,F \gets \textsc{Choose}(T)$
        \State $A_1,\dots,A_{n_p} \gets \textsc{Partition}(\bm{X})$
        \State $N_p \gets N_p + n_p - 1$
        \For{$1 \leq l \leq n_p$}
            \State $n_c \gets \min(\floor*{p|\bm{X}|},|A_{i} \cap \bm{X}|)$
            \State $n_s \gets\floor*{n(1-p)/N_p n_c} + 1$
            \State $A' \gets A_{l} \cap A$
            \State Let $\bm{x}'_1,\dots,\bm{x}'_{n'}$ be a relabelling of $A' \cap \bm{X}$ such that $g(\bm{x}'_1) \geq \dots \geq g(\bm{x}'_{n'})$.
            \State $b \gets g(\bm{x}'_{n_c})$
            \State $ F' \gets \{\bm{x} \in \R^d: g(\bm{x}) \geq b\}$
            \State $f \gets f(\bm{x}) \one_{A'}(\bm{x}) \one_{F'}(\bm{x})$
            \For{$ 1 \leq i \leq n_c $}
                \State $\bm{x}_{i,1} \gets \bm{x}'_{i}$
                \For{$ 2 \leq j \leq n_s $}
                    \State $\bm{x}_{i,j} \gets \textsc{Step}_{f}(\bm{x}_{i,j-1})$
                \EndFor
            \EndFor
            \State $\bm{X}' \gets ((\bm{x}_{i,j})_{i=1}^{n_c})_{j=1}^{n_s}$
            \State Make $(\bm{X}',A',F')$ child of $(\bm{X},A,F)$
        \EndFor
    \EndWhile
    \State \textbf{return} $T$
\EndProcedure
\end{algorithmic}
\end{algorithm}

\subsection{Estimating the Probability of Failure}\label{sec:estimating failure probabilities}

The tree object $T$ returned by \ac{NSuS} can be used to estimate any failure probability $P_F$ defined by a threshold $b$ and failure region $F$. In particular, $b$ is not required to be the critical threshold. The estimation of failure probabilities under \ac{NSuS} is done as follows. Firstly, a trimming procedure is carried out on $T$ at each depth level, starting from 0 and increasing. Note that each node in $T$ has an associated intermediate failure region, which in turn has an associated intermediate threshold. For each node at the current depth, consider the children of that node. If any of the children have an associated threshold larger than $b$, then delete all of the children from $T$ and also all of the children's descendants. Secondly, number the leaves of the resulting trimmed tree $1,\ldots, N_p$, and for the $i$th leaf, number the nodes on the unique path from the root to the leaf, $(\bm{X}^i_0,A^i_0,F^i_0), \ldots, (\bm{X}^i_m,A^i_m,F^i_m)$. The sequence of levels $\bm{X}^i_0, \ldots, \bm{X}^i_m $ together with the nested sequence of sets $A^i_0 \cap F^i_0,\ldots,A^i_m \cap F^i_m,A^i_m \cap F $ define a \ac{SuS} estimator $\hat{P}_i$ for $\prob(A^i_m \cap F)$. Since 
\begin{equation}\label{eq:total probability}
    P_F = \prob(F) = \sum_{i=1}^{N_p}  \prob(A^i_m \cap F),
\end{equation}
it is sensible to suggest
\begin{equation}\label{eq:nsus estimator}
    \hat{P}_F = \sum_{i=1}^{N_p}  \hat{P}_i
\end{equation}
as an estimator for $P_F$. Three assumptions are made in the following discussion in order to simplify the analysis of the statistical properties of $\hat{P}_F$: (i) samples generated by different chains are uncorrelated through the indicator function; (ii) intermediate failure thresholds and partitions are chosen a priori rather than dynamically and (iii) the size of all the levels is a constant $n$.

It is shown in \ref{appendix_a} that $\hat{P}_F$ is asymptotically unbiased and consistent by Proposition \ref{prop:unbias} and Proposition \ref{prop:consist} respectively. Let the size of a \ac{SuS} estimator be the number of estimators that make up its product. For any $i$ and $j$, $\hat{P}_i$ and $\hat{P}_j$ can be rewritten as $\hat{P}_i = \hat{P}_{ij} \hat{P}_a $, $\hat{P}_j = \hat{P}_{ij} \hat{P}_b $, where $\hat{P}_{ij},\hat{P}_a, \hat{P}_b$ are \ac{SuS} estimators and $\hat{P}_{ij}$ has maximum possible size.  It is allowed that $\hat{P}_{ij}=1$. $\hat{P}_{ij}$ can be thought of as a common root estimator. Let  $\delta_{ij}$ denote the \ac{CoV} of $\hat{P}_{ij}$, which can be estimated by a standard SuS \ac{CoV} estimator $\hat{\delta}_{ij}$, and $\hat{w}_{ij} = \hat{P}_i\hat{P}_j / \sum_{l,k=1}^{N_p} \hat{P}_l\hat{P}_k$. An estimator for $\delta$, the \ac{CoV} of \ac{NSuS}, justified by Proposition \ref{prop:consist} and Proposition \ref{prop:cov_est}, is therefore given by
\begin{equation}\label{eq:nsus cov}
    \delta \approx \hat{\delta} = \sqrt{\sum^{N_p}_{i,j=1} \hat{w}_{ij}\hat{\delta}_{ij}^2}.
\end{equation}
It is also possible to use the output of \ac{NSuS} to produce $X \sim f|F$ as follows. First, sample $X_i \sim f|A^i_m \cap F$ for $1 \leq i \leq N_p$ using a Markov chain and then randomly pick one of the $X_i$ weighted by $\prob(A^i_m \cap F)$. Proposition \ref{prop:sample} shows that the resulting sample will have the required distribution.

\ifSubfilesClassLoaded{%
    \newpage 
    \bibliography{references}%
}

\end{document}

\section{The Partitioner}\label{sec:the partitioner}

The fundamental component of \ac{NSuS} is the partitioner. This section provides details on its construction and properties.

\subsection{Niching}\label{sec:niching}

In \ac{EMO}, niching methods are used to maintain individuals in multiple separate neighbourhoods of local maxima, called niches \cite{xiaodongliNichingNichingParameters2010}. The central idea underlying \ac{NSuS} is that each separate high density area of a failure region, such as the neighbourhoods of a design points, can be interpreted as a niche. If \ac{NSuS} is able to consistently explore all the niches of a failure region, the resulting probability of failure estimator will have a comparatively desirable distribution. It follows that it should be possible to adapt existing niching methods to construct effective partitioners. Specifically, if a partitioner returns a partition where the constituent sets correspond to the niches of a failure region, \ac{NSuS} is guaranteed to populate all the important areas of that failure region.

For the purposes of this paper, a niching method is more specifically defined as a process which assigns labels to a set of samples in a level. Samples with the same label belong to the same niche. A general strategy for designing a partitioner for \ac{NSuS} is to first apply a niching method to a level, and then train a classifier on the labeled samples to produce a partition of the input space. In the following two subsections, the \ac{HVG} partitioner is introduced. This is the partitioner used in the examples of this paper. The \ac{HVG} partitioner is a modular framework consisting of three steps: (i) the \ac{HVG} construction, (ii) community detection on the \ac{HVG}, (iii) and classification. The process is depicted in Figure \ref{fig:pipeline}. The first two steps combine to make a new niching method that is particularly well suited to high dimensional problems. 

\begin{figure}
\makebox[\textwidth]{
\begin{tikzpicture}[node distance=3.8cm]
    \node (start) [startstop,fill=color2] {Level};
    \node (pro1) [process, right of=start,fill=color3] {Hill valley graph};
    \node (pro2) [process, right of=pro1,fill=color3] {Community detection};
    \node (pro3) [process, right of=pro2,fill=color3] {Classification};
    \node (stop) [startstop, right of=pro3,fill=color2] {Partition};

    \draw [arrow] (start) -- (pro1);
    \draw [arrow] (pro1) -- (pro2);
    \draw [arrow] (pro2) -- (pro3);
    \draw [arrow] (pro3) -- (stop);

    \draw [decorate,decoration={brace,amplitude=0.6cm,raise=0.8cm}]
        (start) -- (stop) node[midway,yshift=1.8cm]{Hill valley graph partitioner};
\end{tikzpicture}
}
\vspace*{+0mm}
\caption{The pipeline of constituent steps of the hill valley graph partitioner.}
\label{fig:pipeline}
\end{figure}

Before introducing the \ac{HVG} partitioner, the definition of niche needs to be refined for the structural reliability context. Notice that there are no local maxima in the piecewise linear function example depicted in Figure \ref{fig:sus piecewise}. However, the samples can be clearly sorted into two groups based on the gradient of the performance function at their position: those with gradient vectors pointing in the positive $x_1$ direction, and those  with gradient vectors pointing in the positive $x_2$ direction. The concept of convexity can be used to establish a definition of a niche that is useful in such situations. The standard definition of convexity is binary: a set $F \subset \R^d$ is either convex or not convex. When this set is endowed with a probability distribution, the way a failure region is, it becomes natural to measure the degree of convexity of $F$. The following definition formalises this idea. 
\begin{definition}[Convexity Measure]\label{def:convexity measure}
Let $F \subset \R^d$ be a set endowed with a distribution with probability density function $f$. Let $\bm{x}$, $\bm{y} \in F$. The convexity measure of $F$ is given as:
\begin{equation}\label{eq:convexity measure}
    \mathcal{C}_f(F) = \iint_{F} f(\bm{x}) f(\bm{y})\chi_{F}(\bm{x},\bm{y}) d\bm{x}d\bm{y},
\end{equation}
where
\begin{equation}\label{eq:convexity indicator}
    \chi_{F}(\bm{x},\bm{y}) = 
    \begin{cases}
        1 & \text{if } \{t\bm{x} + (1-t)\bm{y}: 0 \leq t \leq 1 \} \subset F,\\
        0 & \text{otherwise.}
    \end{cases}
\end{equation}
\end{definition}
It follows immediately from the definition that $0 \leq \mathcal{C}_f(F) \leq 1$. If a failure region is convex, its convexity measure is equal to $1$. The following heuristic is useful for designing a partitioner: the lower the convexity measure of a failure region, the more likely ergodicity problems will arise when a Markov chain is exploring it. This assertion becomes intuitive when the extreme case is considered. Suppose a failure region consists of two disconnected sets of equal probability density and that are extremely far apart. Such a failure region will have a low convexity measure and cause ergodicty problems. That is, if a chain starts in one of the disconnected sets, it has a low probability of ever reaching the other set. This concept can be used to more suitably define a niche in the reliability analysis context: a niche is a subset of a failure region with relatively high convexity. 

\subsection{Hill Valley Graph}\label{sec:hill valley graph}

Given two samples in a level, say $\bm{x}$ and $\bm{y}$, the problem is to decide whether they should belong to the same niche. Suppose that for all failure regions that contained $\bm{x}$ and $\bm{y}$, they also contained the line segment between $\bm{x}$ and $\bm{y}$. According to the aforementioned heuristic, it would make sense that $\bm{x}$ and $\bm{y}$ belong to the same niche, since this would increase the convexity measure of the relevant failure regions that intersect with this niche. The converse is also true: if for some failure region the line segment between $\bm{x}$ and $\bm{y}$ was not contained within the failure region, then the samples should be separate niches, otherwise this would decrease the convexity measure. This approach to deciding if two samples belong to the same niche is known as a hill valley test in \ac{EMO} literature \cite{ursemMultinationalEvolutionaryAlgorithms1999}. The idea is that if two samples are on separate hills of the objective function, with a valley between them, they should be assigned to two different niches. There are three major ways in which niching methods that employ a hill valley test can vary: the choice of sample pairs to test for a valley, the manner in which the tests are conducted and given the results of the tests, how to decide the niches. The following describes the \ac{HVG} partitioner approach to each of these considerations.

To begin, the \ac{HVG} partitioner randomly selects $n_g$ graph samples from the input level, where $n_g$ is a user defined parameter called the graph size. The random selection is done in a specific manner in order to provide greater coverage of the input space. Let chains that start from the same seed be called a chain group. A simplifying assumption is that samples from the same chain group belong to the same niche. Thus, the sample selection procedure first randomly chooses a number of chain groups with uniform probability and without replacement, and then chooses a random representative of each chain group again with uniform probability. Next, all possible pairs of graph samples are hill valley tested. This process results in a graph, where the samples are vertices which are adjacent if no valley is found between them. A community of a graph is a set of vertices which is internally densely connected and sparsely connected to the other vertices in the graph. Due to the manner in which the graph has been constructed, vertices in the same community correspond to samples in the same niche, and so the problem of deciding niches has been transformed into community detection on a graph. It should be noted that this approach avoids utilising the Euclidean metric, which is commonly employed by niching methods with hill valley tests. This has been done so that the resulting partitioner can work well in high dimensions. Now the \ac{HVG} is formally defined.
\begin{definition}[Hill Valley Graph]\label{def:hill valley graph}
Let $\bm{x}_1, \dots, \bm{x}_{n_g}$ denote randomly chosen graph samples from a level. For any pair of samples in a level, the largest failure region that contains them both can be defined:
\begin{equation}\label{eq:largest failure}
    F_{ij}= \{ \bm{x} \in \R^d: g(\bm{x}) \geq \min(g(\bm{x}_i),g(\bm{x}_j))\}.
\end{equation}
The adjacency matrix of the \ac{HVG} for $1 \leq i,j \leq n_g$ is given as
\begin{equation}\label{eq:adjacency}
    A_{ij}= \chi_{F_{ij}}(\bm{x}_i,\bm{x}_j).
\end{equation}
\end{definition}

\begin{figure}
    \centering
    \begin{subfigure}[b]{0.475\textwidth}
        \centering
        \includegraphics[scale=0.35]{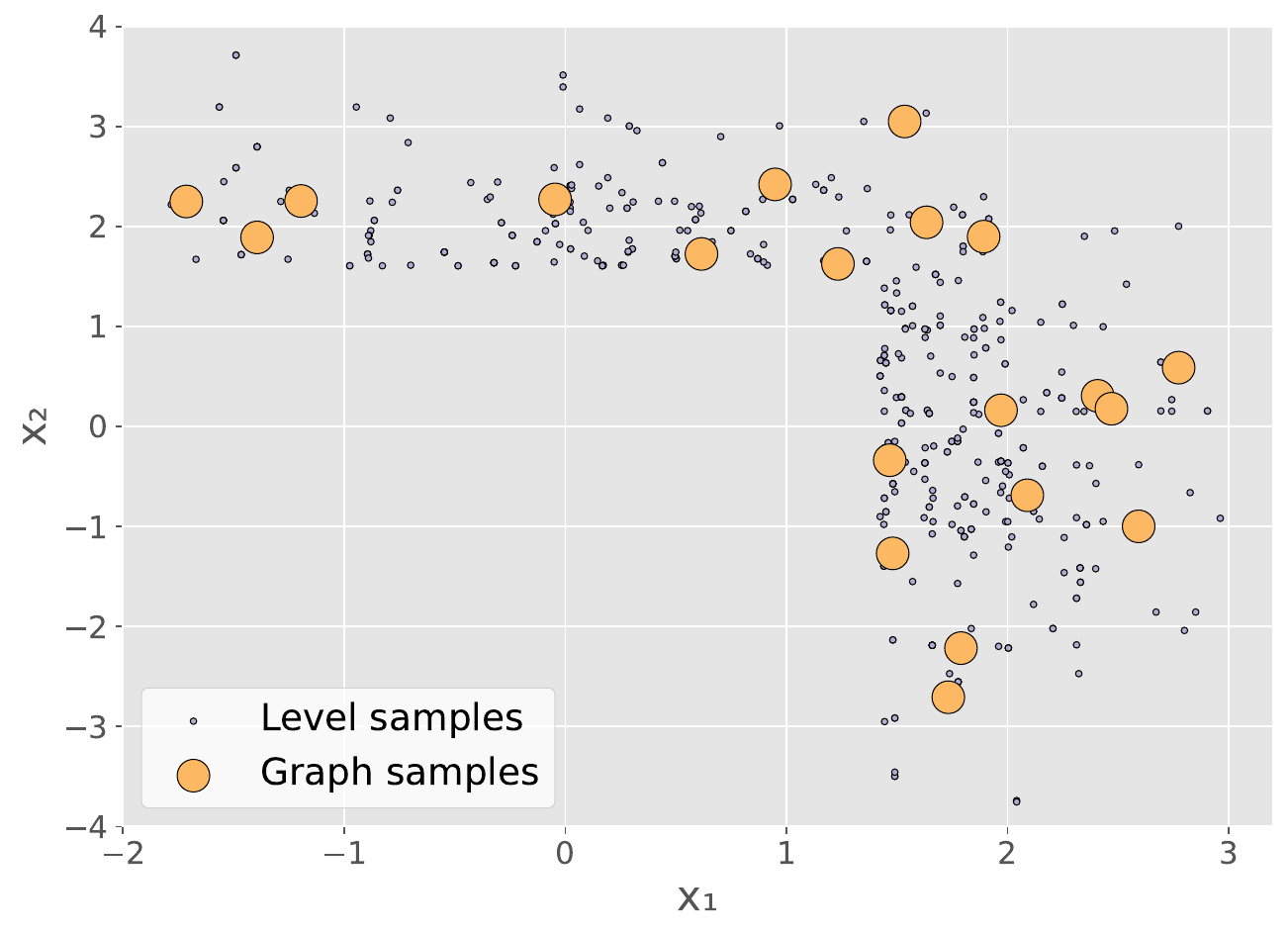}
        \caption{Selecting graph samples}
        \label{fig:graph_samples}
    \end{subfigure}
    \hfill
    \begin{subfigure}[b]{0.475\textwidth}
        \centering
        \includegraphics[scale=0.35]{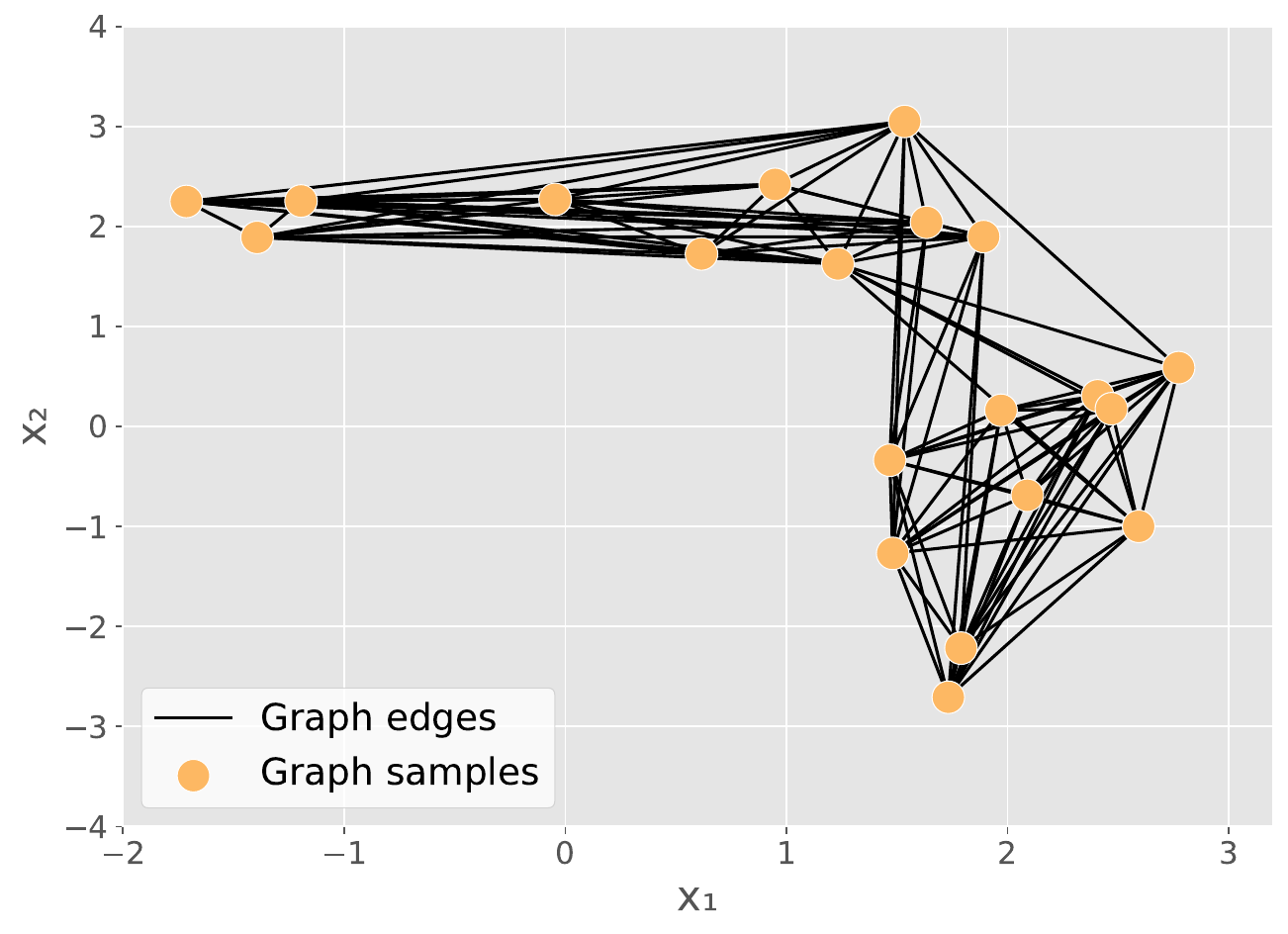}
        \caption{The hill valley graph}
        \label{fig:hvg}
    \end{subfigure}
    \caption{The construction of the hill valley graph on a level defined with the piecewise linear function. First the graph samples are randomly selected from the level, and then the adjacency matrix is calculated.}
    \label{fig:construct hvg}
\end{figure}

Without explicit access to the performance function it is impossible to compute the adjacency matrix of the \ac{HVG}, and so it must be approximated. Hill valley tests are schemes that search for valleys by evaluating the objective function at points on the line connecting two samples. It is common for hill valley tests to use many evaluations, which can be computationally expensive. For this reason, the \ac{HVG} partitioner only evaluates the performance function at the midpoint of the two samples being tested. This approach has a relatively high chance of missing a valley, but the hope is that the community detection algorithm will be able to correct these mistakes using the additional information in the adjacency matrix. Further computational savings can be made by using the open hypercube search history based test introduced in \cite{liHistoryBasedTopologicalSpeciation2015}. For any two samples, $\bm{x} = (x_1,\dots,x_d) $ and $\bm{y}= (y_1,\dots,y_d)$, there exists an open hypercube exists between them,
\begin{equation}\label{eq:hypercube}
    H(\bm{x},\bm{y})= \{e=(e_1,\dots,e_d) : \min(x_i,y_i) < e_i < \max(x_i,y_i), \forall 1 \leq i \leq d\}.
\end{equation}
During a run of \ac{NSuS}, all evaluations of the performance function can be stored in a search history, $\mathcal{S}$. An approximate hill valley test can be performed without evaluating the performance function by checking that all points that are in the intersection, $ H(\bm{x},\bm{y}) \cap \mathcal{S} $, are also in the largest failure region that contains both $\bm{x}$ and $\bm{y}$. This method only practically works in low dimensions, since in high dimensions the intersection will nearly always be empty. Formally combining these approaches gives the approximate \ac{HVG} adjacency matrix,
\begin{equation}\label{eq:approx adjacency matrix}
    \tilde{A}_{ij} =
        \begin{cases}
            \one_{F_{ij}} \left[ H(\bm{x}_i,\bm{x}_j) \cap \mathcal{S}\right] \quad &\text{if } H(\bm{x}_i,\bm{x}_j) \cap \mathcal{S} \neq \emptyset,   \\  
            \one_{F_{ij}}\left[\left(\bm{x}_i + \bm{x}_j\right)/2 \right] \quad &\text{otherwise.}
        \end{cases}
\end{equation}
When $\one_{F_{ij}}$ acts on a set in the above definition, all members of the set must lie in ${F_{ij}}$ in order for a $1$ to be returned. The number of performance function evaluations required to compute the approximate adjacency matrix is bounded above by the number of edges in a complete graph, that is $n_g(n_g -1)/2$. For low  dimensional problems, very few additional evaluations will be required, whereas for high dimensional problems the upper bound will likely be attained. It is a nice feature of the \ac{HVG} partitioner that it automatically and smoothly demands more computational resources as the dimension of the problem increases. The choice of  graph size will be investigated in Section \ref{sec:numerical examples}. When \ac{NSuS} branches, the graph size is split evenly amongst the branches to avoid the computational complexity rapidly rising. The process of constructing the \ac{HVG} is depicted in Figure \ref{fig:construct hvg}. Note that the example input level is level 1 of a \ac{SuS} run on the piecewise linear function, shown in Figure \ref{fig:pwl_3d_sus_2}.

\subsection{Community detection and classification}\label{sec:community detection and classification}

Once the \ac{HVG} has been constructed, community detection algorithms can be used to label the samples. As previously mentioned, the \ac{HVG} partitioner is a modular framework in the sense that any community detection algorithm could be used. One important consideration when deciding between community detection algorithms is the parameters they require the user to define. Some community detection algorithms, such as Louvain Community Detection algorithm \cite{blondelFastUnfoldingCommunities2008}, require the user to specify a resolution parameter. The higher the resolution parameter the more attention that is paid to fine-grained details of the graph and so smaller communities are encouraged. The lower the resolution parameter the more the algorithm focuses on large generalised structures of the graph resulting in larger communities. Despite the fact that $1$ is often used a default value for resolution, there is still the potential for the performance of \ac{NSuS} to be sensitive to such a parameter. Other options, like the Fluid Communities algorithm \cite{paresFluidCommunitiesCompetitive2018}, require the user to a priori specify the number of communities to be identified. If that number is set as $1$, \ac{NSuS} will never branch and so becomes identical to \ac{SuS}. However, if that parameter is set as some number greater than $1$, \ac{NSuS} will branch at every level even if multiple niches do not exist. Branching unnecessarily in this way will adversely affect the efficiency of the algorithm.

With the above in mind, the \ac{ALP} algorithm \cite{raghavanLinearTimeAlgorithm2007} is an attractive option since it does not require the user to specify any parameters. For this reason, ALP is used as the core of the community detection algorithm in the examples of this paper. The details of ALP are given in \ref{appendix_b}, Section \ref{sec:alp}. \ac{ALP} is a stochastic algorithm, and with different starting seeds can produce different community labellings for the same input graph. To account for this, the \ac{HVG} partitioner runs the algorithm 100 times with different random seeds. If more than $50\%$ of labellings consist of only one community, the entire input space is returned as the partition and \ac{NSuS} does not branch. Some of the community labellings may have a high number of communities. This can be problematic for the specific implementation of \ac{NSuS} used in this paper since when a branching occurs, the computational resources are split evenly amongst the branches. If there are many branches, due to many communities, it may be difficult to produce a good statistical estimator since too few computational resources are allocated to each branch. For this reason it is useful for the user to be able to specify a maximum number of branches that \ac{NSuS} can split into, denoted  $N_{p}^{\text{max}}$. To account for this, any of the 100 community labellings that contain more than $N_{p}^{\text{max}} - N_p + 1$ communities must be edited. Given an unacceptable labelling, two communities are randomly chosen and combined into one community, repeatedly, until there are only $N_{p}^{\text{max}} - N_p + 1$ communities remaining.

For each unique community labelling in the graph space, there exists a corresponding classification problem in the input space. That is, each vertex can transfer its community label to the corresponding graph sample where it now acts as a class label. Any labelling with only one community is not considered. The next step of the \ac{HVG} partitioner is to produce potential partitions by training classifiers on the newly labelled sets of graph samples. There are of course many different possible classification algorithms and preprocessing pipelines that could be used at this step. For the numerical examples in this paper, \ac{LSVC} is used. The main reason for this is that \ac{LSVC} is well suited to high-dimensional problems, which are common in the context in which SuS is applied. Details of \ac{LSVC} are given in \ref{appendix_b}, Section \ref{sec:lsvm}. All the samples in the level, not just the randomly selected graph samples, are used for data preprocessing. First, the level is normalised such that it has 0 mean and unit variance, and then the principal components of the scaled level are computed and the graph samples are transformed into the principal component basis.

\begin{figure}
    \centering
    \begin{subfigure}[b]{0.475\textwidth}
        \centering
        \includegraphics[scale=0.35]{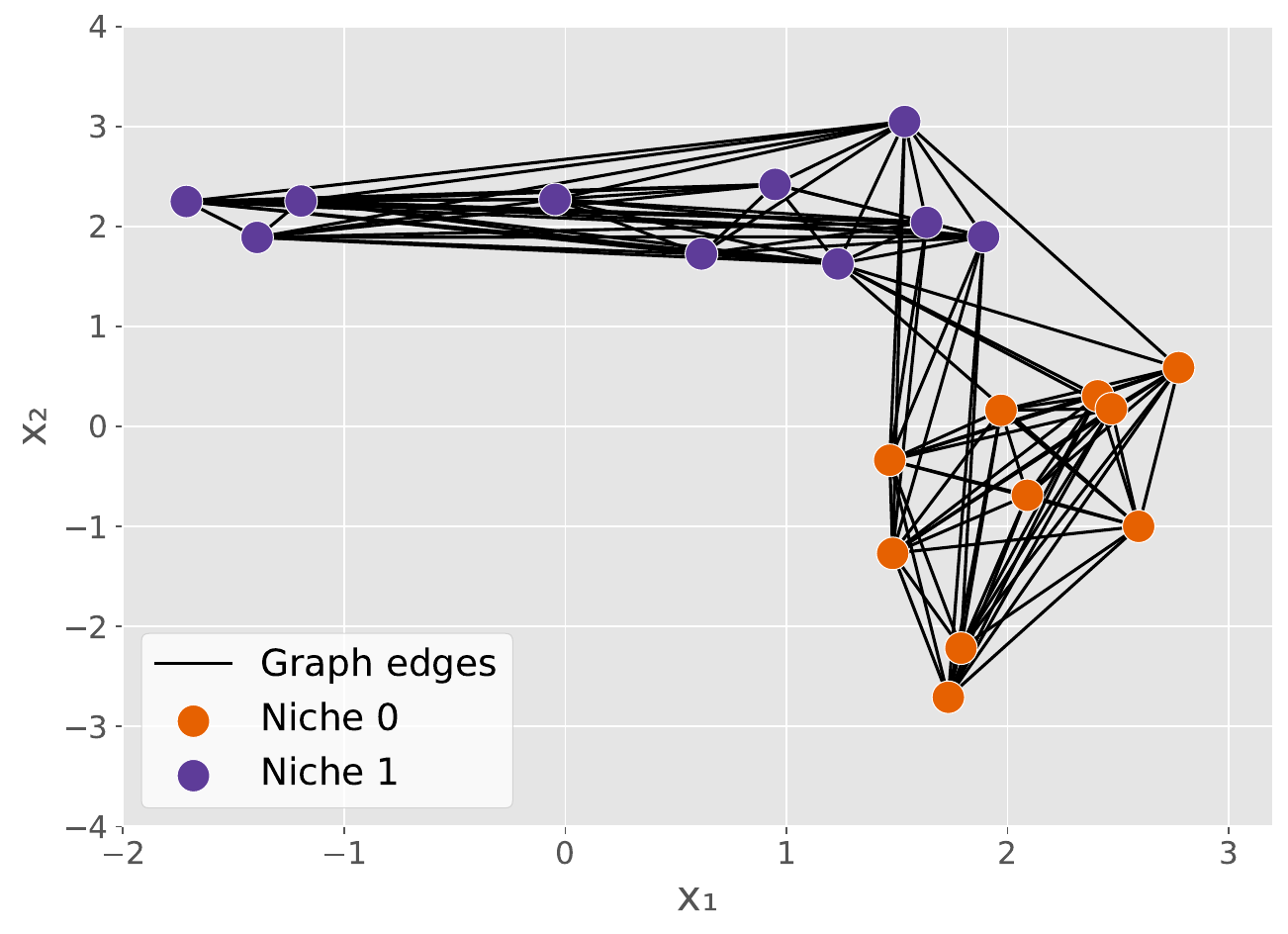}
        \caption{Community detection on hill valley graph}
        \label{fig:classifier}
    \end{subfigure}
    \hfill
    \begin{subfigure}[b]{0.475\textwidth}
        \centering
        \includegraphics[scale=0.35]{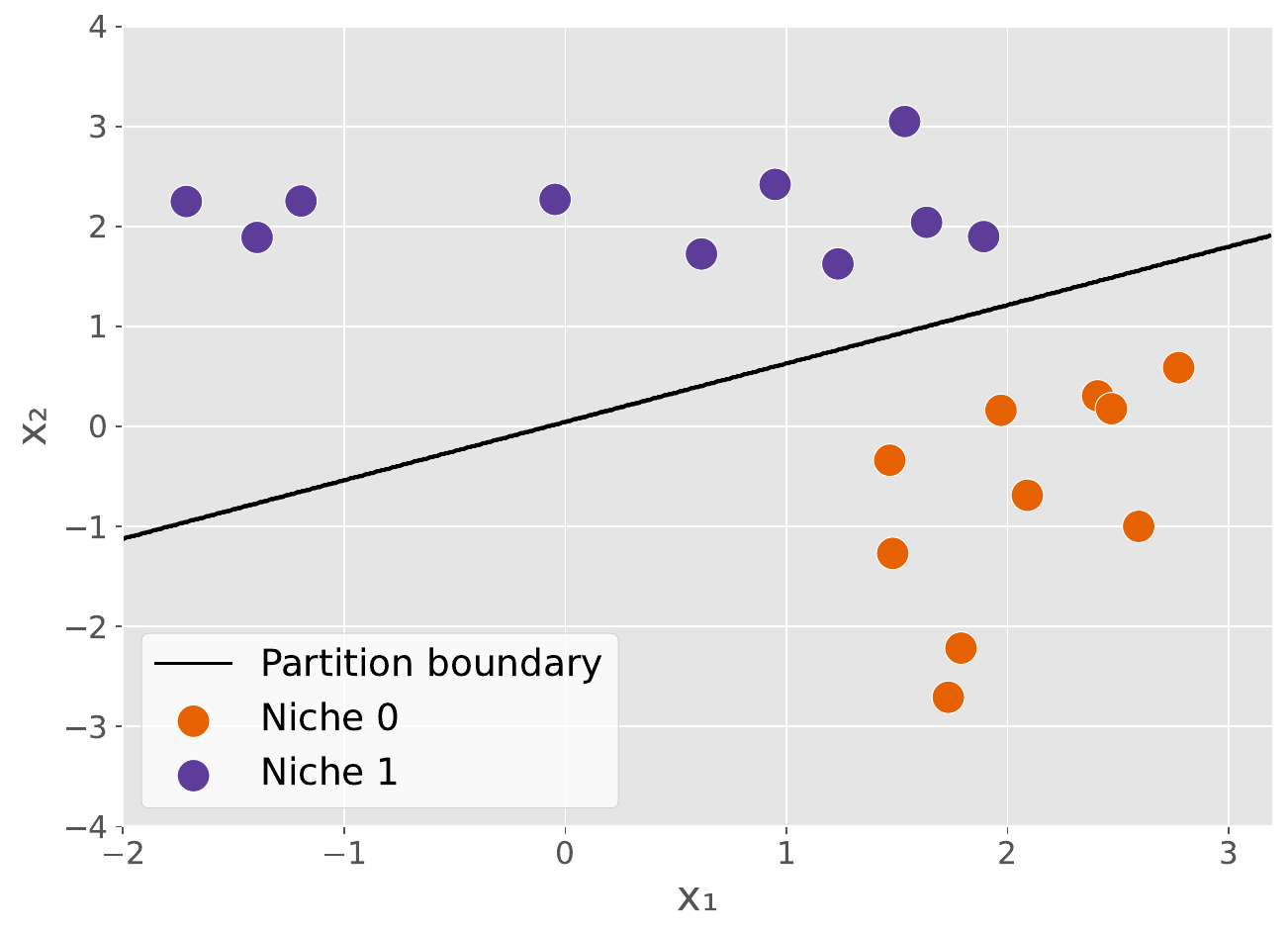}
        \caption{Partition defined by classifier}
        \label{fig:classifier_original}
    \end{subfigure}
    \caption{Community detection is applied to a hill valley graph and two niches are found. A classifier is then trained on the labelled samples in the input space. The final result is a partition of the input space.}
    \label{fig:community detection and classifier}
\end{figure}

The final step of the \ac{HVG} partitioner is to decide between the partitions produced by the classifiers. The natural approach is to evaluate the classifiers using a classification metric. The numerical examples in this paper compare the true labels to the predicted labels using the balanced accuracy score. The balanced accuracy score has been chosen since it is able to deal with both imbalanced and multi-label classification problems. Details of the balanced accuracy score are given in \ref{appendix_b}, Section \ref{sec:bas}. The partition with the highest balanced accuracy score is the final output of the \ac{HVG} partitioner. Figure \ref{fig:community detection and classifier} depicts the community detection and classification steps of the \ac{HVG} partitioner. This is a continuation of the piecewise linear function example shown in Figure \ref{fig:construct hvg}. Note that the community detection step is shown in the input space for visualisation purposes only. In reality, it is crucial that the community detection step takes place in the graph space, without any access to the coordinate information, since this allows the \ac{HVG} partitioner to be effective in high dimensions.

\ifSubfilesClassLoaded{%
    \newpage 
    \bibliography{references}%
}

\end{document}

\section{Numerical Examples}\label{sec:numerical examples}

The examples presented in this section have topologies that prevent \ac{SuS} from consistently and fully exploring their failure regions. In such cases, the resulting probability of failure estimators typically have bimodal distributions. The modes are representative of the \ac{SuS} runs that do or do not populate the neighbourhoods of the design points, which correspond to overestimation and underestimation. In the most dramatic examples, the modes can differ by several orders of magnitude from each other, and from the true probability of failure. That is, even though the mean of distribution could be a good estimate of the true probability of failure, there is very little density in the neighbourhood of the target. There also exists an asymmetry between the two cases. A gross underestimation may not be that different to the true probability of failure in absolute terms, but can have a very different scale, and the opposite is true for overestimation. The performance of a Monte Carlo based reliability method is usually evaluated using the mean and the \ac{CoV} of the probability of failure estimator. Both of these metrics are reported in this section, but due to the aforementioned reasons, they are not sufficient for assessing the efficiency of the tested algorithms. For each example, a degeneracy indicator is defined to decide if a simulation run has populated all of the neighbourhoods of the design points. Using this, the estimates are labeled as ``degenerate'' or ``non-degenerate'', and the resulting distributions are summarised separately, as well as jointly. Perhaps most importantly, the proportion of degenerate runs is also given, denoted as ``weight'' in the tables.

The numerical experiments were conducted using 100 simulations of both \ac{SuS} and \ac{NSuS}. For each experiment, \ac{SuS} has a fixed level size, and \ac{NSuS} has a fixed level size, graph size, and maximum number of branches. The probability of failure estimates vary within an experiment due to different random seeds. The computational cost is measured as the average number of performance function evaluations, since in practical scenarios these evaluations typically dominate the algorithm run time. However, it should be noted that the \ac{HVG} partitioner does incur additional computational cost when performing community detection and training classifiers. For \ac{NSuS} the average number of performance function evaluations is reported as $x +y$, where $x$ is the number of evaluations used to create the levels, and $y$ is the number evaluations required to construct the \ac{HVG}. For the purpose of reproducibility, the Python code and data used in this section are available on GitHub at \url{https://github.com/HughKinnear/nsus_paper}. The code uses the Scikit-learn LSVC implementation \cite{feurerPracticalAutomatedMachine2018} and the NetworkX ALP implementation \cite{SciPyProceedings_11}. 

\subsection{Piecewise Linear Function}\label{sec:plf}

\begin{figure}
    \centering
    \begin{subfigure}[b]{\textwidth}
        \includegraphics[scale=0.7]{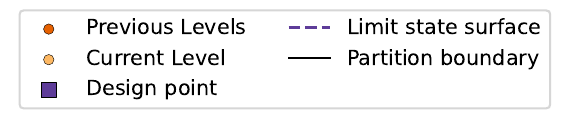}
    \end{subfigure}
    \begin{subfigure}[b]{0.475\textwidth}
        \centering
        \includegraphics[scale=0.55]{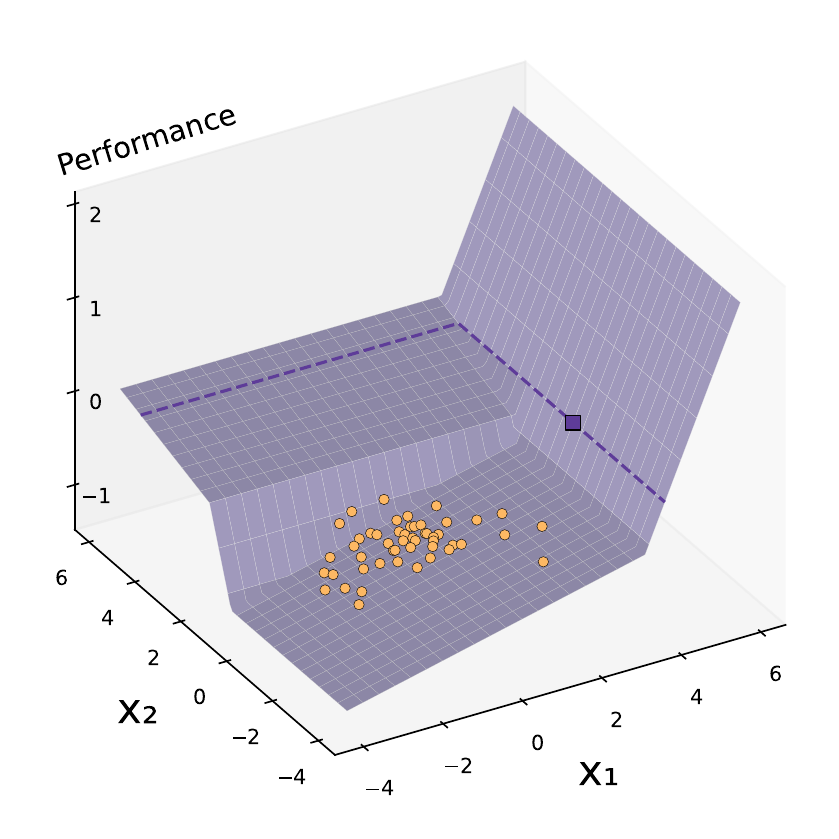}
        \caption{Initial level}
        \label{fig:pwl_3d_nss_1}
    \end{subfigure}
    \begin{subfigure}[b]{0.475\textwidth}
        \centering
        \includegraphics[scale=0.55]{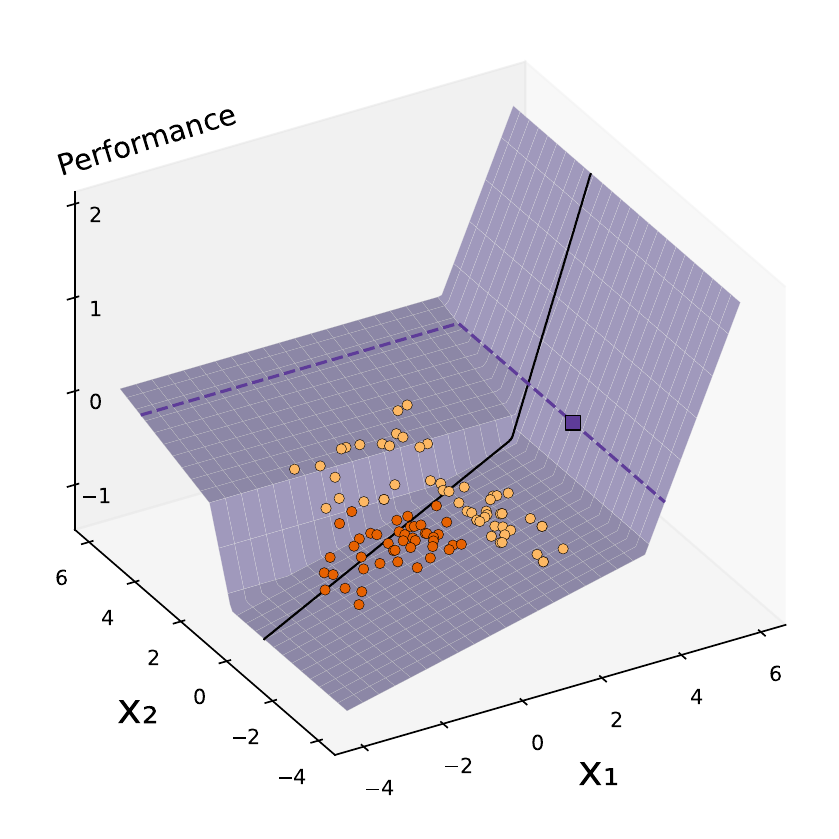}
        \caption{Level $1$}
        \label{fig:pwl_3d_nss_2}
    \end{subfigure}
    \begin{subfigure}[b]{0.475\textwidth}
        \centering
        \includegraphics[scale=0.55]{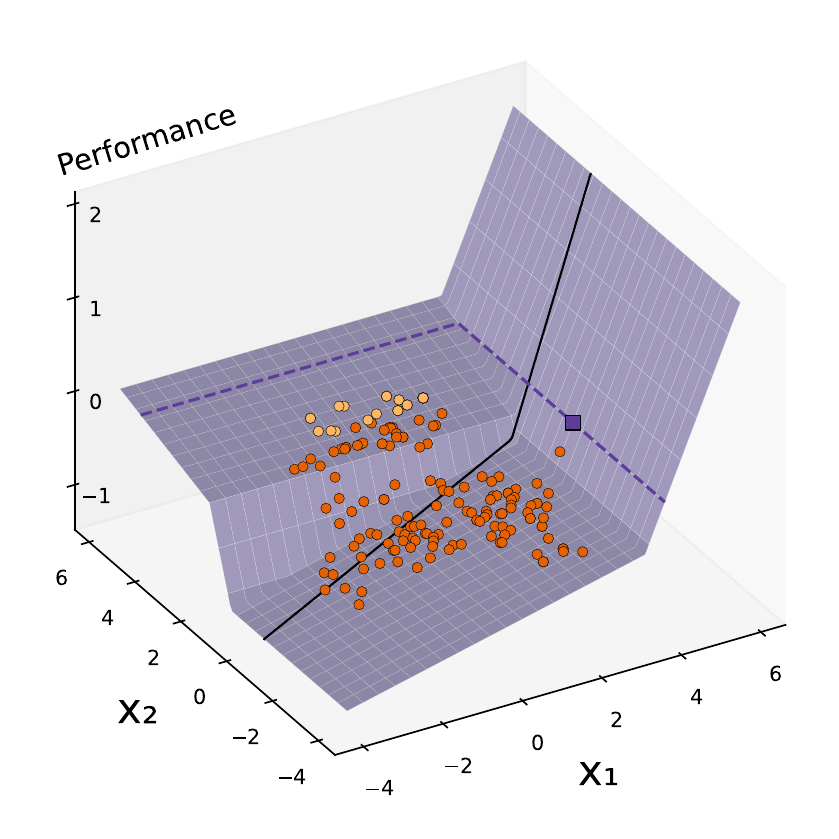}
        \caption{Level $4$}
        \label{fig:pwl_3d_nss_5}
    \end{subfigure}
    \begin{subfigure}[b]{0.475\textwidth}
        \centering
        \includegraphics[scale=0.55]{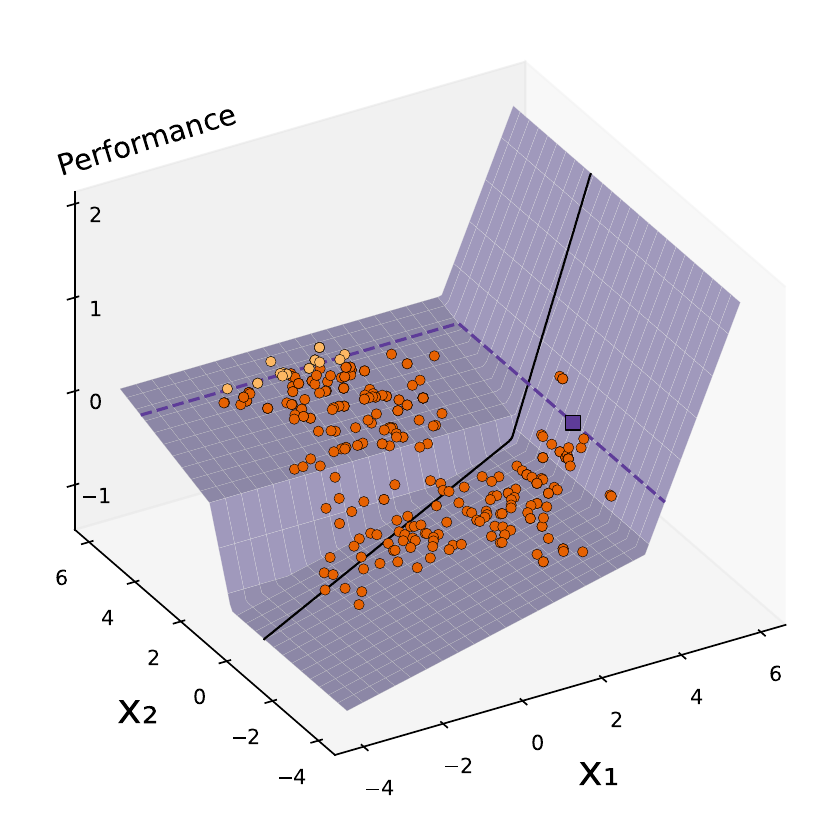}
        \caption{Level $11$}
        \label{fig:pwl_3d_nss_12}
    \end{subfigure}    
    \caption{NSuS running on the piecewise linear function. For clarity, only every $10^{th}$ sample has been plotted. NSuS is able to populate the neighbourhood of the design point by partitioning the input space and exploring each set of the partition.}
    \label{fig:pwl_nss}
\end{figure}

The first example revisits the piecewise linear function, that is $g$ as defined in equation \ref{eq:piecewise linear} in Section \ref{sec:niching subset simulation}. In that section, Figure \ref{fig:sus piecewise} depicted a degenerate \ac{SuS} run on the piecewise linear function. As a counterpart, Figure \ref{fig:pwl_nss} shows a non-degenerate run using \ac{NSuS}. It can be seen that \ac{NSuS} starts identically to \ac{SuS} at the initial level, since no niches are detected. However, at level $1$, the \ac{HVG} partitioner returns a partition with two sets. The entire internal process of the \ac{HVG} partitioner acting on level $1$ was depicted in Figures \ref{fig:construct hvg} and \ref{fig:community detection and classifier}. This partition separates the currently high performing samples that will travel in the positive $x_2$ direction from the currently low performing samples that will eventually travel towards the design point. \ac{NSuS} is then started again in each set of the partition, allowing them both to be fully explored. The end result is that the neighbourhood of the design point is populated, which in turn will lead to a good estimate of the probability of failure.

The results of the numerical experiments for the piecewise linear function are reported in Table \ref{table:pwl_2}. The degeneracy indicator checks for samples in $\{\bm{x} \in \R^2: x_1\geq 4,x_2 \leq 2\} \cap F$. In this example, the mean probability of failure of the degenerate runs is a gross underestimation of the true reference value, and can be considered useless. For this reason, the most useful point of comparison is the non-degenerate weight, which is much higher for \ac{NSuS} than \ac{SuS}. Note that whilst the mean probability for the joint \ac{SuS} estimator is a decent estimate, very few individual runs will actually give estimates in this range. \ac{NSuS} also has a significantly lower \ac{CoV} in this example. The average number of performance evaluations is lower for the joint \ac{NSuS} estimator compared to the joint \ac{SuS} estimator, since \ac{NSuS}, due to the allocation scheme, devotes less computational resources to the less important niche than \ac{SuS} does. Note also the negligible amount of performance function evaluations required to construct the \ac{HVG}. This is due to the piecewise linear function being only 2-dimensional. Another benefit of \ac{NSuS} is that even in the cases where \ac{NSuS} produces a degenerate run, it has the potential to alert the user to the existence of potentially challenging geometry. That is, a \ac{HVG} with any missing edges could be cause for concern, even if \ac{NSuS} is not able to successfully maintain the relevant niches. In contrast, when \ac{SuS} has a degenerate run, it is indistinguishable from a non-degenerate run.

\begin{figure}
    \centering
    \includegraphics[scale=0.5]{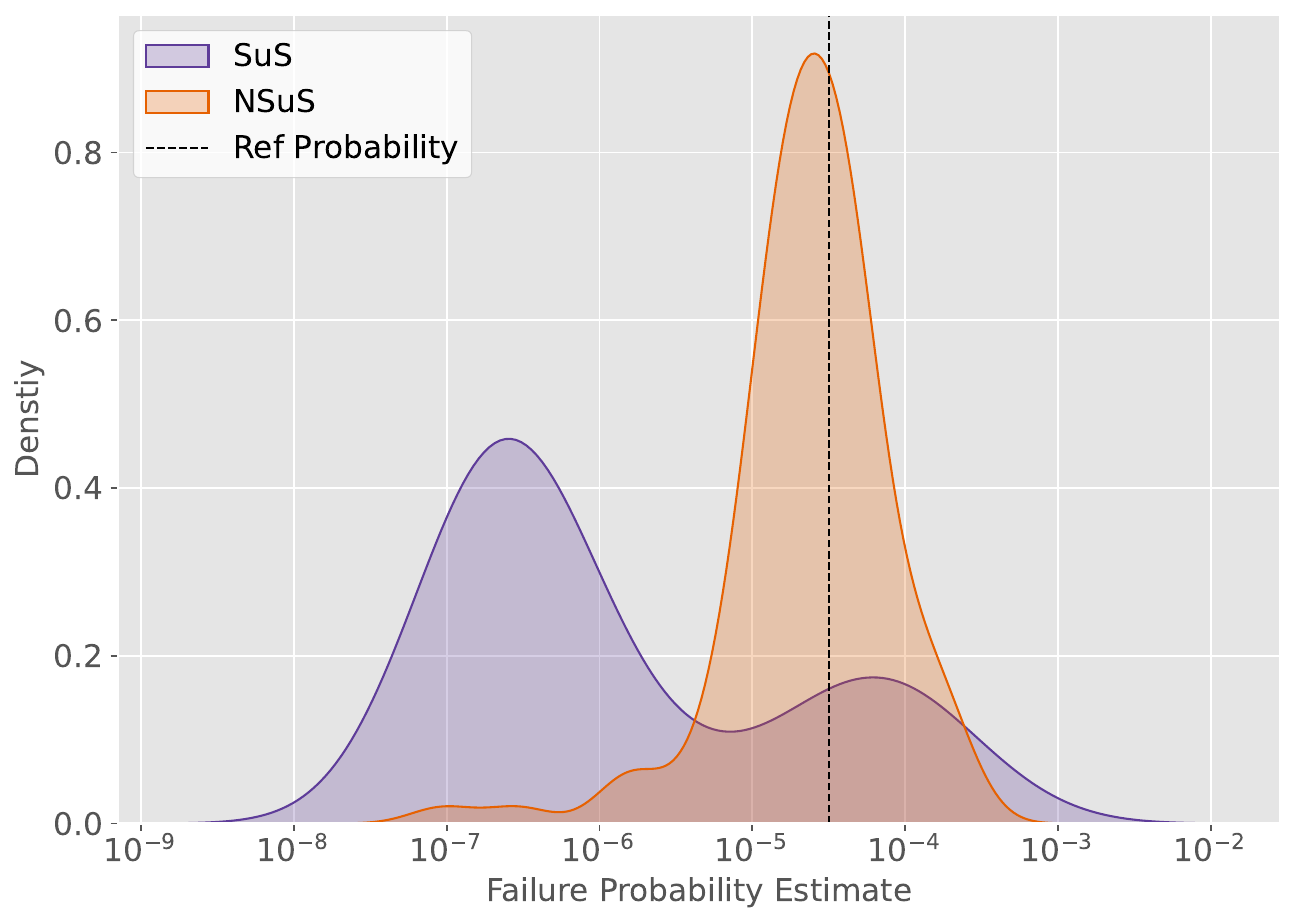}
    \caption{Kernel density estimates of the 100 SuS and NSuS probability failure estimators on the piecewise linear function, compared to a reference probability.}
    \label{fig:nss_kde}
\end{figure}

Figure \ref{fig:nss_kde} shows a kernel density of estimate of both the \ac{SuS} and \ac{NSuS} estimators. A log scale is used, since the estimators vary across orders of magnitude. In general, it can be seen that the \ac{NSuS} has much more density near the reference probability.

\begin{table}[h]
\centering
\begin{tabular}{cccccc}
    \hline
    Method & Estimator & Weight & Mean $P_F$ & CoV $P_F$ & Mean $g$ evals \\
    \hline
    SuS & Joint & 1 & $2.95 \times 10^{-5}$   & 2.61 & 5280 \\
        & Degenerate  & 0.65& $4.26 \times 10^{-7}$  & 1.12& 5819 \\
        & Non-degenerate & 0.35 & $8.36 \times 10^{-5}$   & 1.34 & 4278\\
    \hline
    NSuS & Joint & 1 & $3.96 \times 10^{-5}$   & 1.10 & 4934 + 4 \\
         & Degenerate  & 0.02 & $8.61 \times 10^{-7}$  & 0.89 & 5038 + 1 \\
         & Non-degenerate & 0.98 & $4.04 \times 10^{-5}$   & 1.08 & 4932 + 4\\
    \hline
\end{tabular}
\caption{Piecewise linear function numerical results. Reference probability $3.18 \times 10^{-5}$ estimated with \ac{DMC} estimator with $10^8$ samples. Level size 1000, graph size 15, max branches 2.}
\label{table:pwl_2}
\end{table}

To understand how the input dimension affects \ac{NSuS} performance, a 100-dimensional piecewise linear function is now defined:
\begin{equation}\label{eq:piecewise linear 100}
\begin{aligned}
    & g_{100}(\bm{x}) = g(z_1,z_2), \,\text{where}\\
    & z_1 = \frac{1}{\sqrt{50}}\sum_{i=1}^{50} x_i, \quad z_2 = \frac{1}{\sqrt{50}} \sum_{i=51}^{100} x_i.
\end{aligned}
\end{equation}
The new degeneracy indicator checks for samples in $\{\bm{x} \in \R^{100}: z_1\geq 4,z_2 \leq 2\} \cap F$. Since both $z_1$ and $z_2$ have a standard normal distribution, the 2-dimensional and 100-dimensional performance functions have identical probabilities of failure. They also both exhibit the same challenging geometry. Table \ref{table:pwl_100} shows the results of the numerical experiments for the 100-dimensional piecewise linear function. Compare with the results of the experiments on the 2-dimensional piecewise linear function. \ac{NSuS} has a slightly lower non-degenerate weight. This is likely due to the classifier struggling slightly more in higher dimensions, since the \ac{HVG} construction is independent of dimension. Conversely, \ac{SuS} has a higher non-degenerate rate. The extra dimensions change the manner in which the Markov chains explore the input space, which leads to a lower chance of ergodicity problems. Still, \ac{NSuS} outperforms \ac{SuS} even in higher dimensions. Note that the number of performance function evaluations required to construct the \ac{HVG} is significantly higher than in the low dimensional case. This is because the hypercube between level samples is nearly always empty in 100 dimensions.

\begin{table}[h]
\centering
\begin{tabular}{cccccc}
    \hline
    Method & Estimator & Weight & Mean $P_F$ & CoV $P_F$ & Mean $g$ evals \\
    \hline
    SuS & Joint & 1 & $3.75 \times 10^{-5}$   & 1.34 & 5311 \\
        & Degenerate  & 0.40 & $3.76\times 10^{-7}$  & 0.64 & 6422 \\
        & Non-degenerate & 0.60 & $6.22 \times 10^{-5}$   & 0.84 & 4570 \\
    \hline
    NSuS & Joint & 1 & $4.13 \times 10^{-5}$   & 1.10 & 5154 + 225 \\
         & Degenerate  & 0.05 & $2.93 \times 10^{-7}$  & 0.89 & 6456 + 462 \\
         & Non-degenerate & 0.95 & $4.34 \times 10^{-5}$   & 1.08 & 5085 + 212\\
     \hline
\end{tabular}
\caption{100-dimensional piecewise linear function numerical results. Reference probability $3.18 \times 10^{-5}$ estimated with \ac{DMC} estimator with $10^8$ samples. Level size 1000, graph size 15, max branches 2.}
\label{table:pwl_100}
\end{table}

\begin{figure}
    \centering
    \begin{subfigure}[b]{\textwidth}
        \centering
        \includegraphics[trim={0cm 0.7cm 0cm 0cm},clip, scale=0.7]{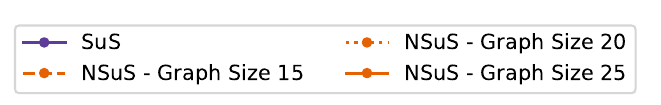}
    \end{subfigure}
    \begin{subfigure}[b]{0.475\textwidth}
        \centering
        \includegraphics[scale=0.35]{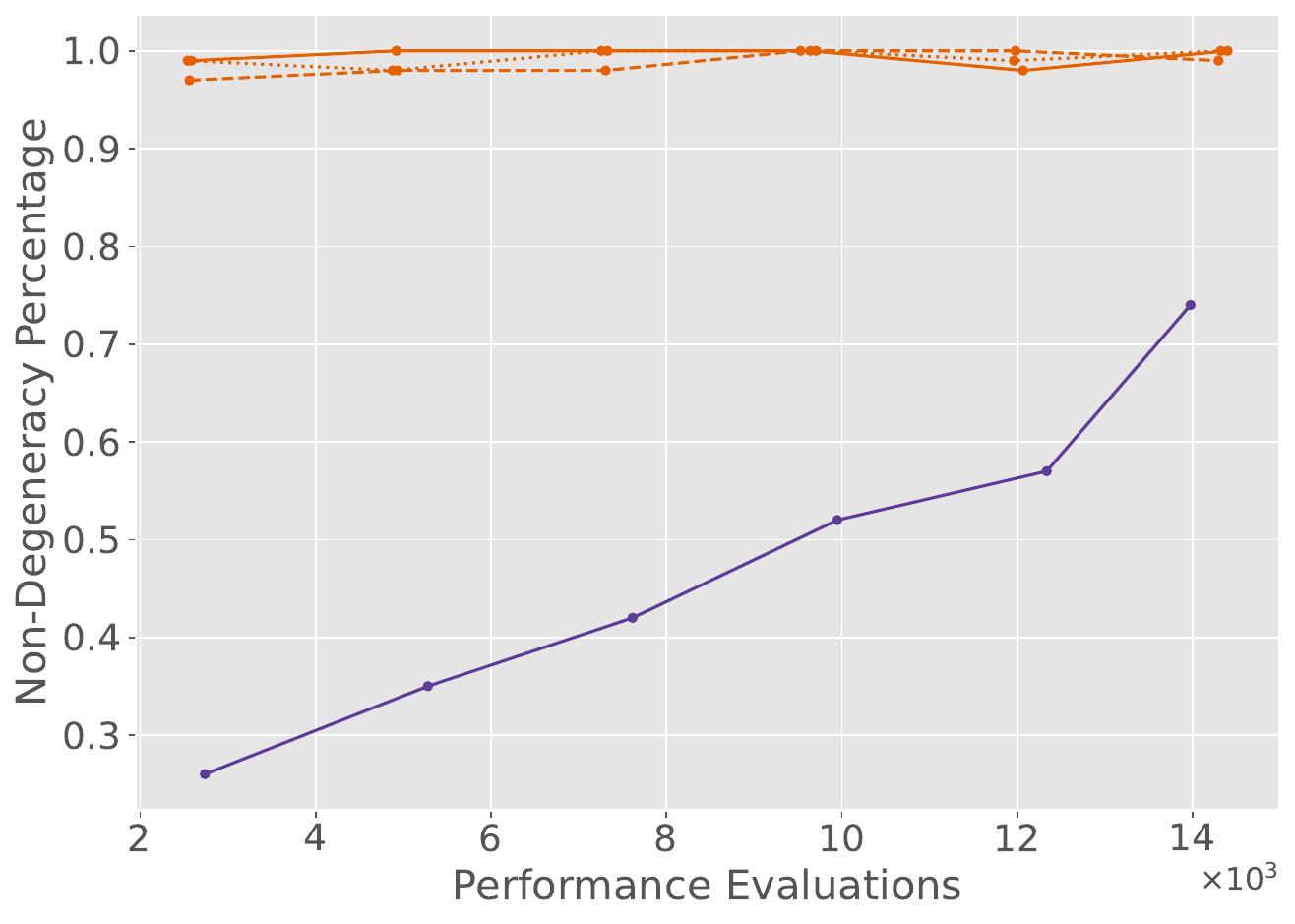}
        \caption{2 dimensions}
        \label{fig:pwl_degen}
    \end{subfigure}
    \hfill
    \begin{subfigure}[b]{0.475\textwidth}
        \centering
        \includegraphics[scale=0.35]{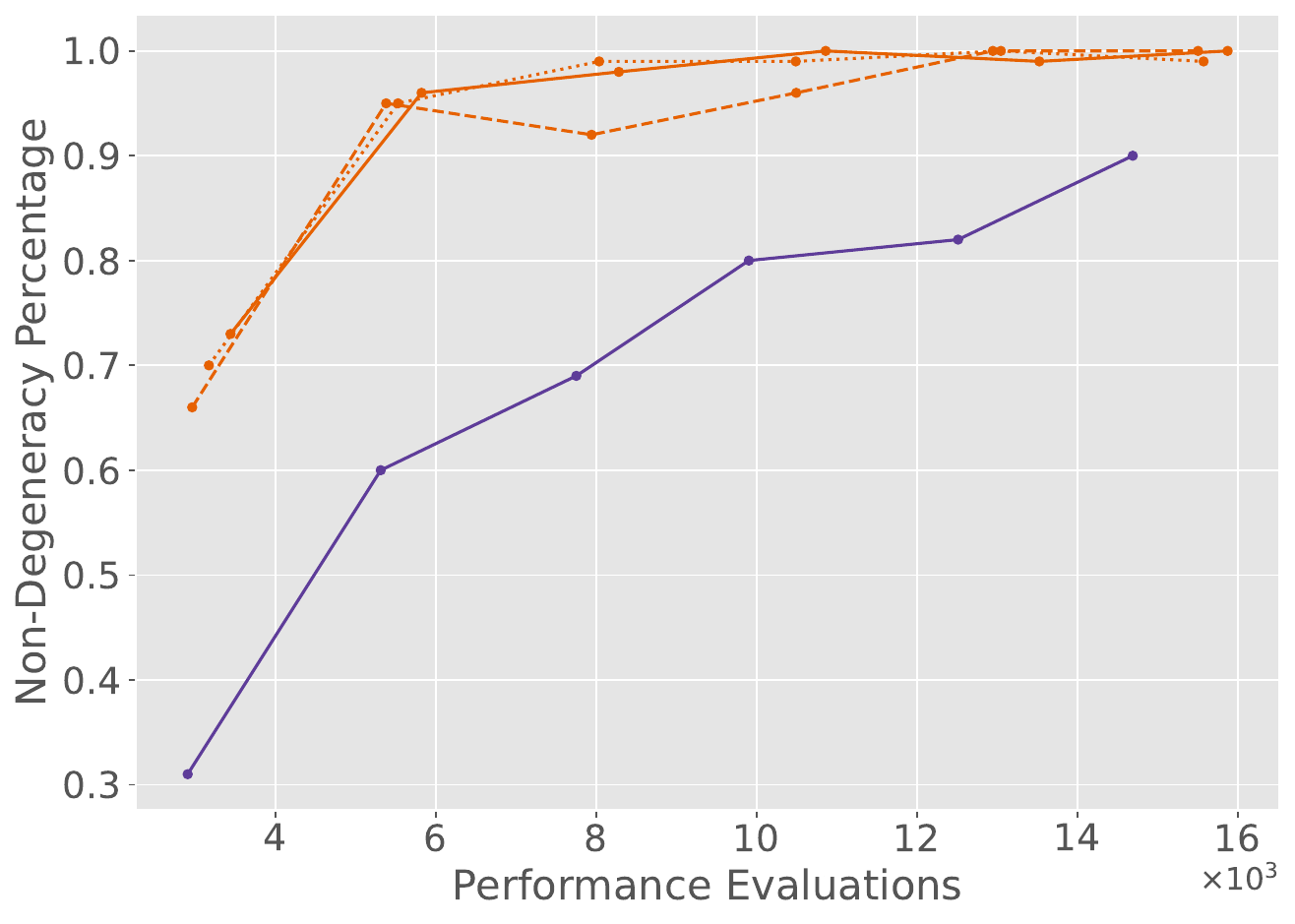}
        \caption{100 dimensions}
        \label{fig:high_pwl_degen}
    \end{subfigure}
    \hfill 
    \caption{Results of \ac{SuS} and \ac{NSuS} numerical experiments on both the 2-dimensional and 100-dimensional piecewise linear function. For both performance functions and algorithms, level sizes of 500, 1000, 1500, 2000, 2500, 3000 are used, represented by scatter points. Linear interpolation has been done between the scatter points. For \ac{NSuS} a range of graph sizes are used. \ac{NSuS} outperforms \ac{SuS} across the range parameter values and computational costs tested in both low and high dimensions.}
    \label{fig:vary_param}
\end{figure}

Finally, Figure \ref{fig:vary_param} explores the effects of varying the level size and graph size in the low and high dimensional case. Experiments were run with both algorithms with level sizes 500, 1000, 1500, 2000, 2500, 3000 and with graph sizes 15, 20, 25 for \ac{NSuS} in both the low and high dimensional cases. It can be seen clearly that \ac{NSuS} outperforms \ac{SuS} across the entire range of parameters tested, in terms of non-degeneracy percentage and number of performance function evaluations. However, it should be noted that the disparity is smaller in higher dimensions for this numerical example.

\subsection{Mixture of Normal Distributions}

This example creates a performance function from a mixture of normal distributions. Let $\varphi(\bm{x};\mu)$ be a 2-dimensional normal probability density function with mean $\mu$ and unit covariance matrix. The performance function is defined as
\begin{equation}\label{eq:mixture}
    g(\bm{x}) = \sum_{i=1}^{4} (w_i \varphi(\bm{x};\mu_i)) - 0.04,
\end{equation}
where
\begin{equation}\label{eq:mix_params}
\begin{bmatrix}
    w_1 \\
    w_2 \\
    w_3 \\
    w_4
\end{bmatrix}
=
\begin{bmatrix}
    0.4 \\
    0.2 \\
    0.2 \\
    0.2
\end{bmatrix}
\, \text{and} \,
\begin{bmatrix}
    \mu_1 \\
    \mu_2 \\
    \mu_3 \\
    \mu_4 
\end{bmatrix}
=
\begin{bmatrix}
    \phantom{-}3 & \phantom{-}3 \\
    \phantom{-}2 & -2 \\
    -2 & \phantom{-}2 \\
    -2 & -2 
\end{bmatrix}.
\end{equation}
The $0.04$ adjustment is only made so that the critical threshold is the conventional value of $0$.

\begin{figure}
    \centering
    \begin{subfigure}[b]{\textwidth}
        \centering
        \includegraphics[trim={0cm 1cm 0cm 0cm},clip, scale=0.7]{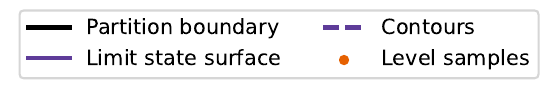}
    \end{subfigure}
    \begin{subfigure}[b]{0.475\textwidth}
        \centering
        \includegraphics[scale=0.48]{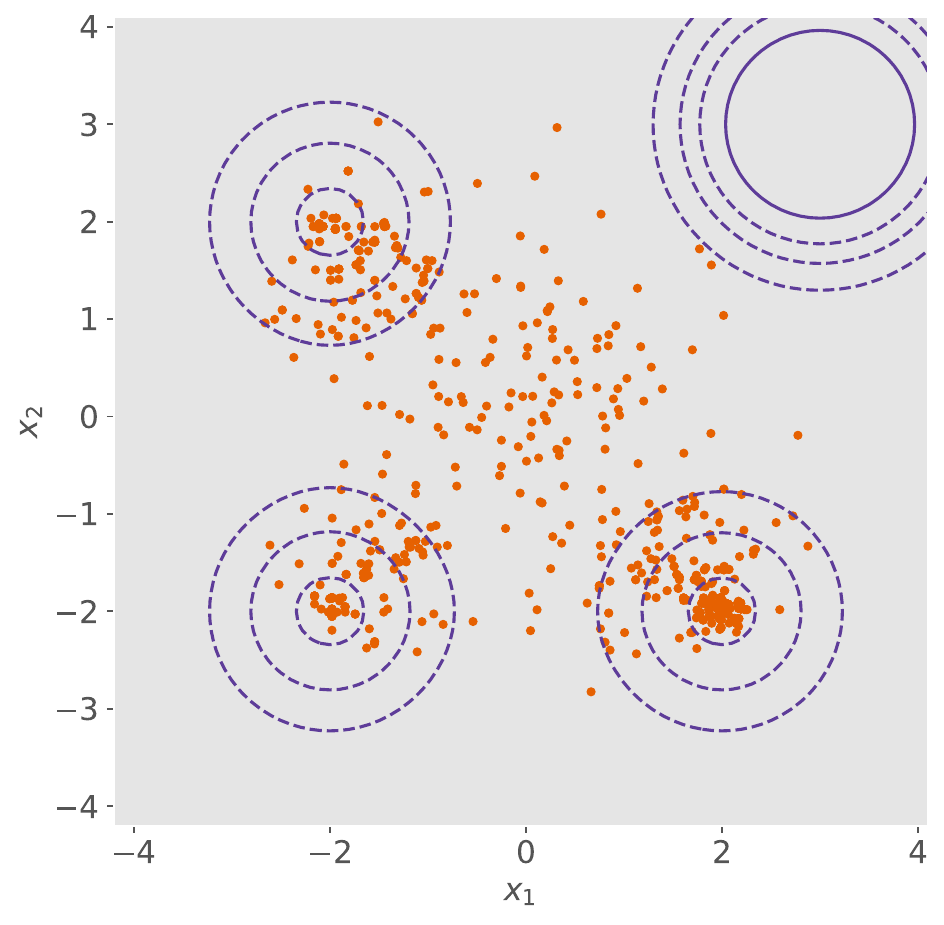}
        \caption{Subset Simulation}
        \label{fig:mix_contour_sus}
    \end{subfigure}
    \begin{subfigure}[b]{0.475\textwidth}
        \centering
        \includegraphics[scale=0.48]{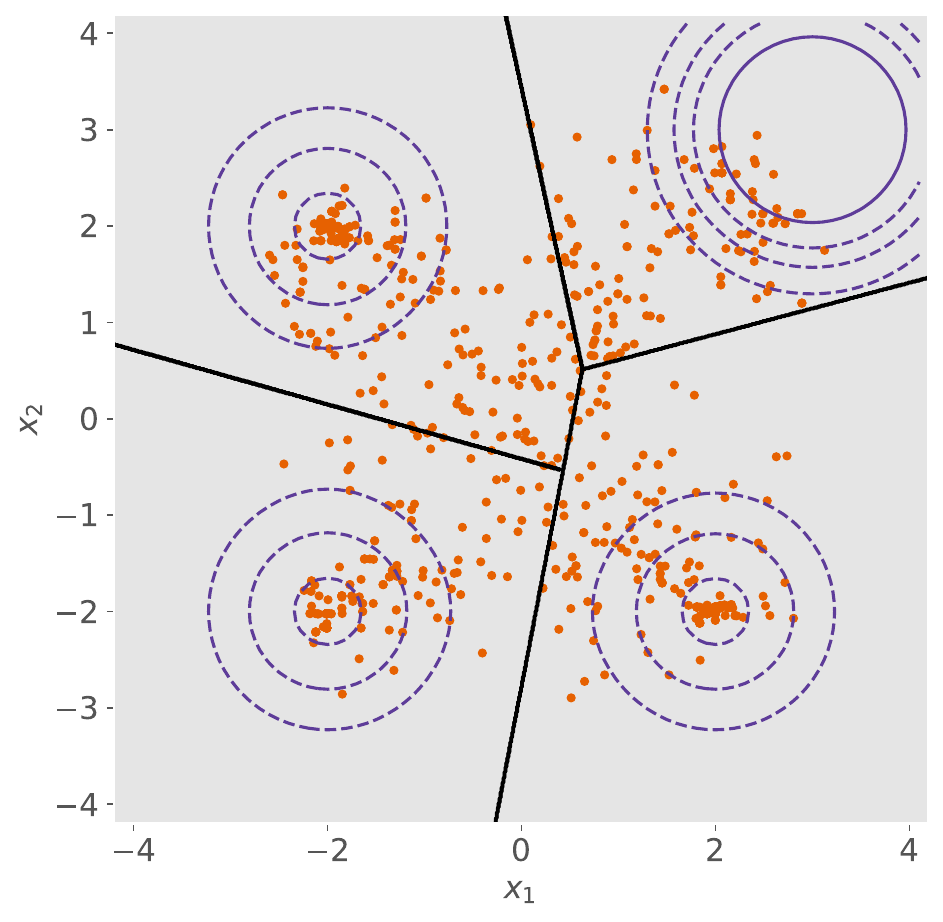}
        \caption{Niching Subset Simulation}
        \label{fig:mix_contour_nss}
    \end{subfigure}
    \hfill 
    \caption{Contour plot of the mixture of normal distributions function. \ac{SuS} gets attracted to the three maxima closest to the origin, and consequently is not able to produce any failure samples. \ac{NSuS} is able create a partition where each maxima gets a corresponding set. Since each set of the partition gets fully explored, \ac{NSuS} is able to produce failure samples.}
    \label{fig:mix_contour}
\end{figure}

Figure \ref{fig:mix_contour} shows runs of \ac{SuS} and \ac{NSuS} on the mixture of normal distributions on top of a contour plot of the performance function. The neighborhood of the three maxima closest to the origin contain none of the failure region, whereas the neighborhood of the maximum furthest from the origin contains the entire failure region. This is problematic for \ac{SuS}, since it can be attracted away from the failure region into the supposedly more promising directions. When this happens, no failure samples are produced and the probability of failure estimator falters in the most dramatic possible way. That is, the estimate for the probability of failure is 0. Conversely, \ac{NSuS} is able to identify all 4 niches and so is able to explore all of them properly. Note that in some cases, not all 4 niches are detected on the initial level. For example, it could be the case that 3 niches are detected on the initial level, and on subsequent levels, one of the niches splits into two.

Table \ref{table:mix} shows the results of the numerical examples for the mixture of normal distributions performance function. The degeneracy indicator checks for samples in $\{\bm{x} \in \R^{2}: x_1\geq 0,x_2 \geq 0\} \cap F$. \ac{NSuS} has a higher non-degeneracy weight, a lower \ac{CoV}, and lower number of performance evaluations for the joint estimator than \ac{SuS}. In this example, the cost of a degenerate run is as high as it could be, which means no failure samples produced at all. Note that \ac{CoV} is not defined for the degenerate estimators since the estimate is 0.

\begin{table}[h]
\centering
\begin{tabular}{cccccc}
    \hline
    Method & Estimator & Weight & Mean $P_F$ & CoV $P_F$ & Mean $g$ evals \\
    \hline
    SuS & Joint & 1 & $2.54 \times 10^{-4}$   & 1.28 & 3549 \\
        & Degenerate  & 0.27 & 0 & - & 5973 \\
        & Non-degenerate & 0.73 & $3.49 \times 10^{-4}$   & 0.96 & 2652 \\
    \hline
    NSuS & Joint & 1 & $2.35 \times 10^{-4}$   & 0.77 & 3018 + 16 \\
         & Degenerate  & 0.01 & 0 & - & 3317 + 15 \\
         & Non-degenerate & 0.99 & $2.37 \times 10^{-4}$   & 0.75 & 3015 + 16\\
     \hline
\end{tabular}
\caption{Mixture of normal distributions numerical results. Reference probability $2.19 \times 10^{-4}$ estimated with \ac{DMC} estimator with $10^8$ samples. Level size 750, graph size 25, max branches 6.}
\label{table:mix}
\end{table}

\subsection{Two-degree-of-freedom Mass Spring System}

\begin{figure}
    \centering
    \includegraphics[scale=0.35]{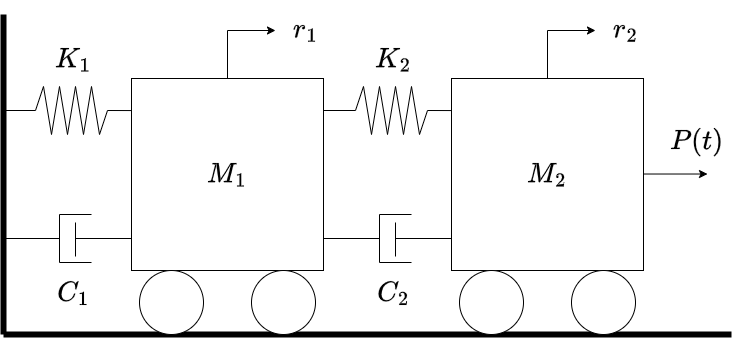}
    \caption{Two-degree-of-freedom mass spring system.}
    \label{fig:two_dof}
\end{figure}

The third example deals with the forced vibration of the \ac{TDOF} mass spring system \cite{sharmaModifiedReplicaExchangebased2023} depicted in Figure \ref{fig:two_dof}. The masses are $M_1 = M_2 = 2000 \text{ kg}$, the modal damping ratios are $\eta_1 = \eta_2 = 0.02$, and forcing function acting on $M_2$ is given as $P(t)=2000\sin(11t)\text{N}$. Let the stiffness parameters $K_1$ and $K_2$ have independent log-normal distributions with mean $2.5 \times 10^5$ and \ac{CoV} 0.2. The standard normal input variables are transformed so that they have the required distribution, $K_1,K_2 = T(x_1,x_2)$. Let $r_1(t;K_1,K_2)$ be the displacement of the first mass at some time, given stiffness constants. The performance function is given as
\begin{equation}\label{eq:two_dof}
    g(x_1,x_2) = \max_{0\leq t \leq20} r_1(t;K_1,K_2) - 0.024.
\end{equation}

\begin{figure}
    \centering
    \begin{subfigure}[b]{\textwidth}
        \centering
        \includegraphics[trim={0cm 1cm 0cm 0cm},clip, scale=0.7]{contour_legend.pdf}
    \end{subfigure}
    \begin{subfigure}[b]{0.475\textwidth}
        \centering
        \includegraphics[scale=0.48]{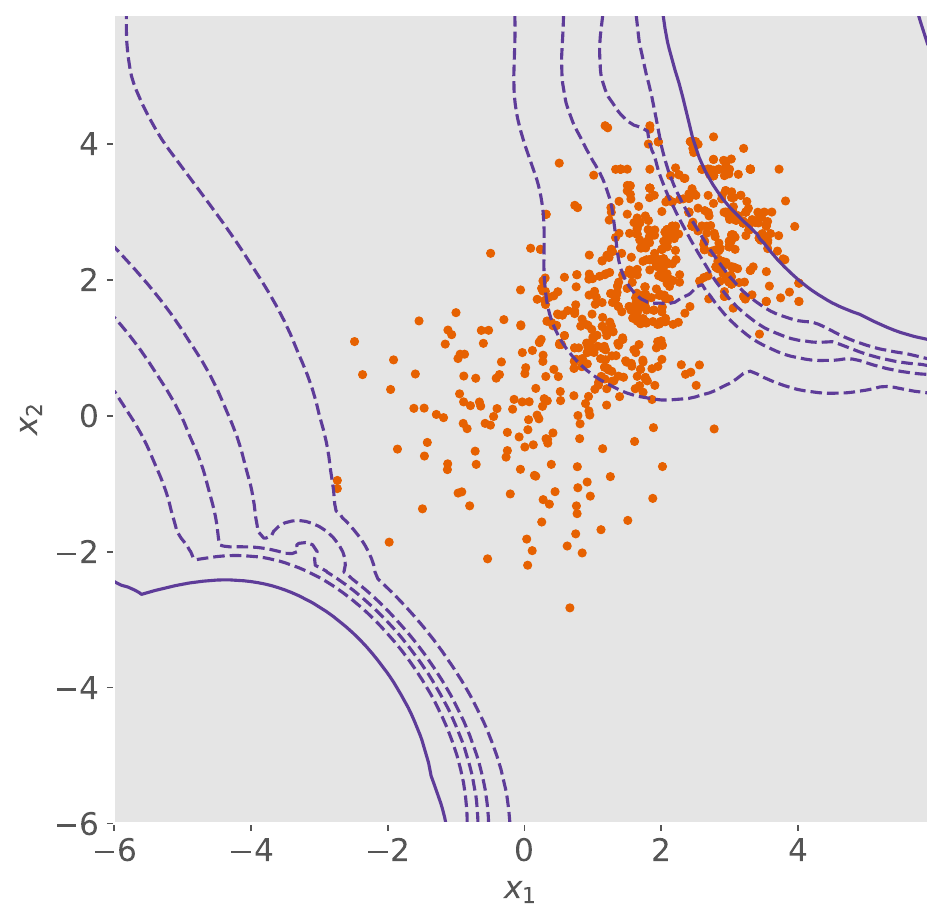}
        \caption{Subset Simulation}
        \label{fig:two_dof_contour_sus}
    \end{subfigure}
    \begin{subfigure}[b]{0.475\textwidth}
        \centering
        \includegraphics[scale=0.48]{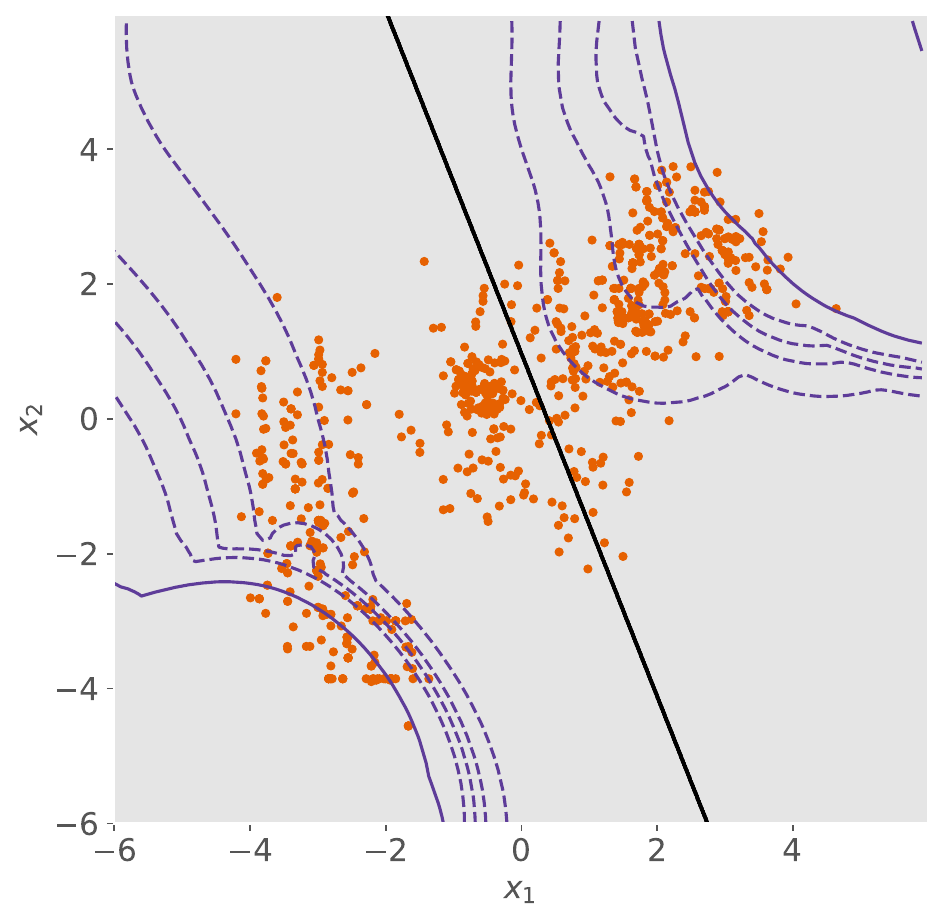}
        \caption{Niching Subset Simulation}
        \label{fig:two_dof_contour_nss}
    \end{subfigure}
    \hfill 
    \caption{Contour plot of the TDOF mass spring performance function. \ac{SuS} is only populates the top right section of the failure region, since the gradient change from the origin is relatively gradual. \ac{NSuS} is able to detect the valley between the two disconnected sets of the failure region and so the partition successfully allows both segments to be populated.}
    \label{fig:two_dof_contour}
\end{figure}

Figure \ref{fig:two_dof_contour} shows runs of \ac{SuS} and \ac{NSuS} on top of contour plots of the performance function. The failure region is split into two disjoint sets, where the bottom left set contributes slightly more to the probability of failure. However, the gradient change on the path towards the top right set is more gradual and so \ac{SuS} is likely to only travel in that direction. The degeneracy indicator for this example checks for samples in $\{\bm{x} \in \R^{2}: x_1\leq 0\} \cap F$. This example is different to the previous two, in that the degenerate runs still give a decent estimate for the probability of failure, since the two disconnected sets of the failure region have similar probability density. \ac{NSuS} is able to detect the valley it starts in at the initial level and so it is able to produce a partition that allows both failure sets to be populated. Table \ref{table:two_dof} reports the results of the numerical results for the mass spring system performance function. \ac{NSuS} has the more desirable estimator since it has a higher non-degenerate weight and much lower \ac{CoV} than the \ac{SuS} estimator.

\begin{table}[h]
\centering
\begin{tabular}{cccccc}
    \hline
    Method & Estimator & Weight & Mean $P_F$ & CoV $P_F$ & Mean $g$ evals \\
    \hline
    SuS & Joint & 1 & $4.15 \times 10^{-5}$   & 1.54 & 3282 \\
        & Degenerate  & 0.45 & $8.28 \times 10^{-6}$  & 0.74 & 3523 \\
        & Non-degenerate & 0.55 & $6.87 \times 10^{-5}$   & 1.11 & 3084 \\
    \hline
    NSuS & Joint & 1 & $2.83 \times 10^{-5}$   & 0.82 & 3145 + 16 \\
         & Degenerate  & 0.08 & $6.30 \times 10^{-6}$  & 0.98 & 3237 + 24 \\
         & Non-degenerate & 0.92 & $3.02 \times 10^{-5}$    & 0.76 & 3137 + 16\\
     \hline
\end{tabular}
\caption{TDOF mass spring system numerical results. Reference probability $2.48 \times 10^{-5}$ estimated with \ac{DMC} estimator with $10^7$ samples. Level size 750, graph size 30, max branches 2.}
\label{table:two_dof}
\end{table}

\subsection{Single-degree-of-freedom Linear Oscillator}

The final numerical example is an adjusted version of the \ac{SDOF} linear oscillator example from \cite{auEstimationSmallFailure2001}. The displacement of the oscillator is measured over 1 second, using a sampling interval of $\Delta t = 0.01s$. This leads to there being $1/\Delta t +1 = 101$ time instants denoted as $t_i$ for $1\leq i \leq 101$. The oscillator is driven by Gaussian white noise at discrete time intervals, defined using the Gaussian inputs: 
\begin{equation}\label{eq:white_noise}
    W(t_i) = \sqrt{\frac{2\pi S}{\Delta t}}x_i,
\end{equation}
where the spectral intensity is $S=1$. The oscillator's equation of motion is
\begin{equation}\label{eq:osc}
    \ddot{r}(t) + 2\zeta\omega\dot{r}(t) + \omega^{2}r(t) = W(t),
\end{equation}
with natural frequency $\omega=7.85$ rad/s, damping ratio $\zeta=0.02$ and initial conditions $r(0)=0,\dot{r}(0)=0$. The performance function measures the distance of the oscillator from the origin at $t_{101}$. Failure is defined to occur when the measured distance is greater than some displacement threshold, which in this case is $1.2$. Formally,
\begin{equation}\label{eq:osc_perf}
    g(\bm{x}) = |r(t_{101};\bm{x})| - 1.2,
\end{equation}
where $r(t;\bm{x})$ is the displacement over time for a given input white noise. There are two distinct regions of the failure domain: the oscillators with a large positive displacement and oscillators with a large negative displacement. Due to the symmetry of this reliability problem, these regions have identical total probability density. The degeneracy indicator in this case checks if both sets of the failure region are populated, that is it checks for samples in both $\{\bm{x} \in \R^{101}: r(t_{101};\bm{x})\leq 1.2\}$ and $\{\bm{x} \in \R^{101}: r(t_{101};\bm{x})\geq 1.2\}$.

\begin{figure}
    \centering
    \begin{subfigure}[b]{\textwidth}
        \centering
        \includegraphics[trim={0cm 1cm 0cm 0cm},clip, scale=0.7]{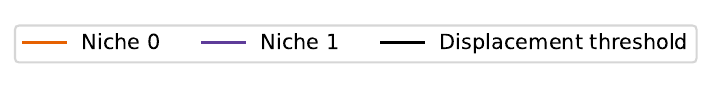}
    \end{subfigure}
    \begin{subfigure}[b]{0.475\textwidth}
        \centering
        \includegraphics[scale=0.32]{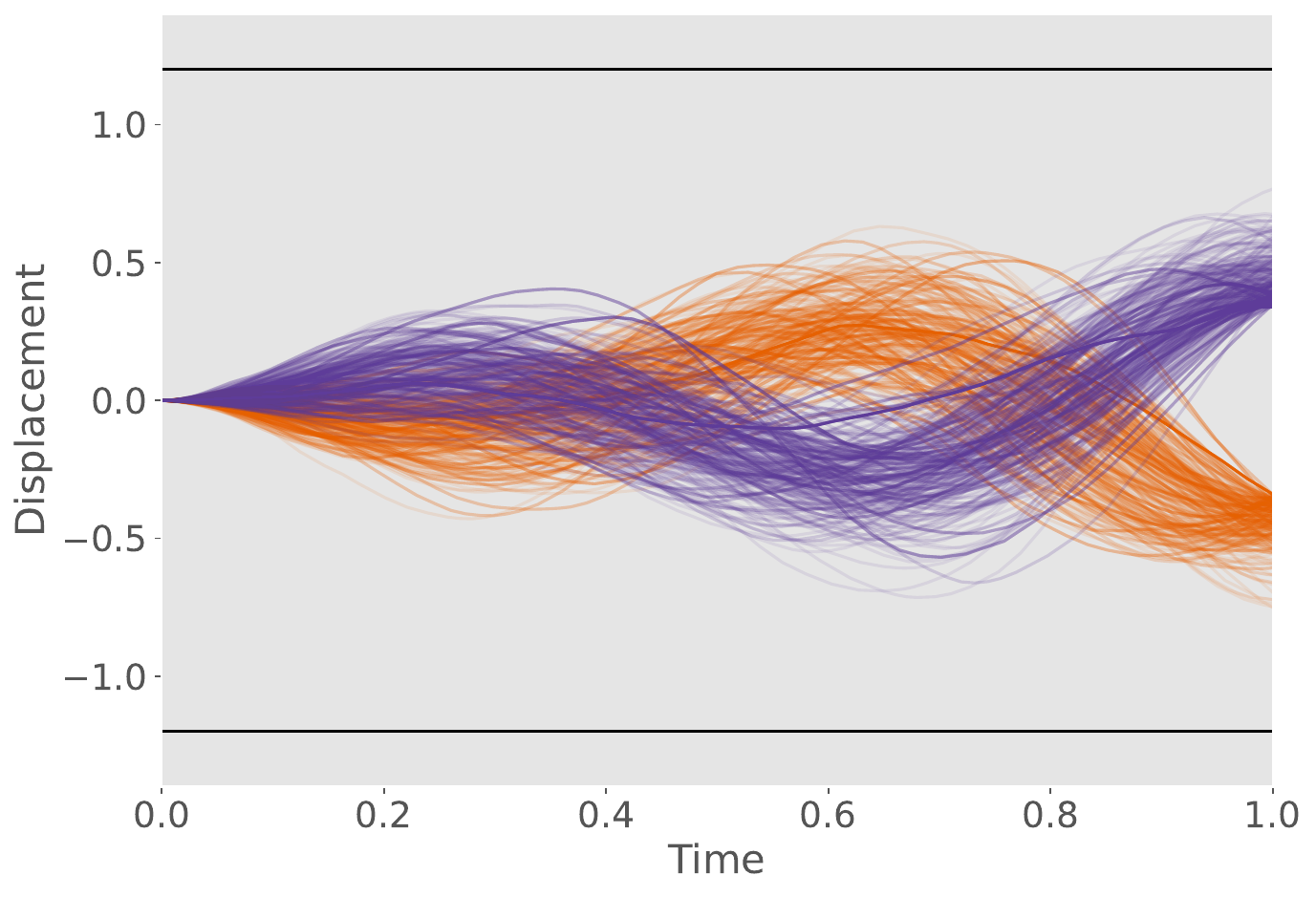}
        \caption{Level 1}
        \label{fig:osc_branch}
    \end{subfigure}
    \begin{subfigure}[b]{0.475\textwidth}
        \centering
        \includegraphics[scale=0.32]{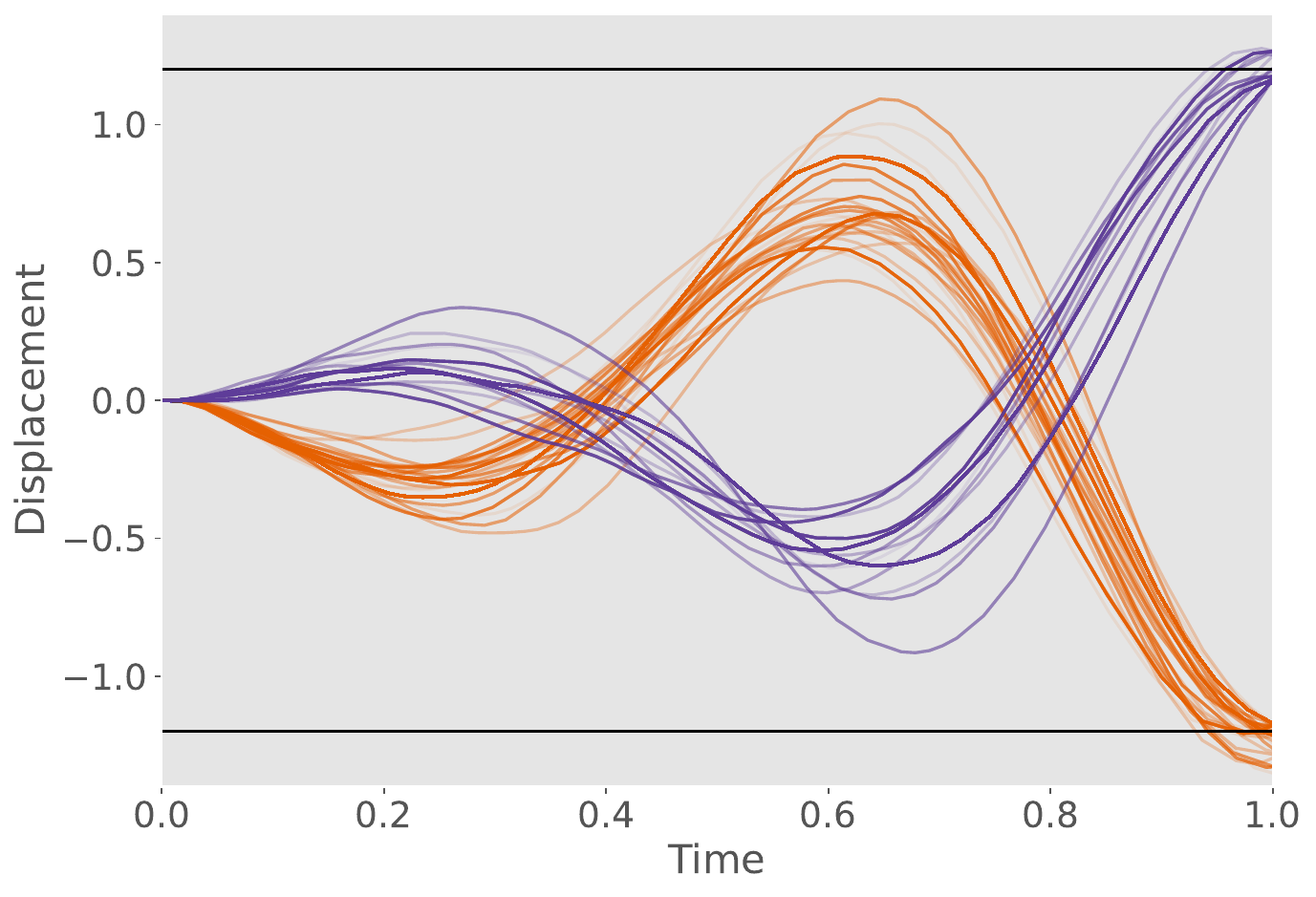}
        \caption{Level 14}
        \label{fig:osc_fail}
    \end{subfigure}
    \hfill 
    \caption{Levels of \ac{NSuS} represented as a displacement over time of the SDOF linear system. On Level 1, \ac{NSuS} partitions the samples into two niches. The niches then continue to develop until they both reach the failure region at level 14.}
    \label{fig:osc_nss}
\end{figure}

Figure \ref{fig:osc_nss} shows the ability of \ac{NSuS} to identify niches in this high dimensional reliability problem. At level 1, \ac{NSuS} is able to identify and split the level into two niches: oscillators that finish with positive displacement, and oscillators that finish with negative displacement. The algorithm continues and is able to maintain both niches, so when the algorithm finishes, failure samples of both type are present. Table \ref{table:osc} reports the results of the numerical experiments for the \ac{SDOF} linear oscillator performance function. Whilst \ac{NSuS} does have a higher non-degenerate weight than \ac{SuS}, in this case this does not appear to have any meaningful positive effect in terms of the distribution of the estimator. This is because the degenerate estimators do give good estimates for the probability of failure. This tends to be the case with performance functions that exhibit some sort of symmetry. However, the additional understanding of the reliability problem that \ac{NSuS} provides can still be useful. If an engineer is attempting to improve a design, knowledge of different failure modes and their relative contribution could be helpful. If the performance function is actually a surrogate for a more computationally expensive performance function, there may be some discrepancy between the model and the target, and points from all niches can be used to validate the model. In the case where a practitioner has many reliability problems that they have reason to believe have similar topology, a deeper understanding of one of the performance functions could simplify the analysis of the rest.

\begin{table}[h]
\centering
\begin{tabular}{cccccc}
    \hline
    Method & Estimator & Weight & Mean $P_F$ & CoV $P_F$ & Mean $g$ evals \\
    \hline
    SuS & Joint & 1 & $3.01 \times 10^{-8}$   & 1.15 & 6214 \\
        & Degenerate  & 0.24 & $2.24 \times 10^{-8}$  & 1.38 & 6290 \\
        & Non-degenerate & 0.76 & $3.26 \times 10^{-8}$   & 1.09 & 6191 \\
    \hline
    NSuS & Joint & 1 & $3.63 \times 10^{-8}$   & 1.07 & 6284 + 198 \\
         & Degenerate  & 0.01 & $7.91 \times 10^{-9}$  & 0 & 6116 + 210 \\
         & Non-degenerate & 0.99 & $3.66 \times 10^{-8}$    & 1.06 & 6286 + 198\\
     \hline
\end{tabular}
\caption{SDOF linear oscillator numerical results. Reference probability $1.46 \times 10^{-8}$ estimated with \ac{SuS} estimator with level size $10^5$. Level size 800, graph size 15, max branches 2.}
\label{table:osc}
\end{table}

\ifSubfilesClassLoaded{%
    \newpage 
    \bibliography{references}%
}

\end{document}

\section{Conclusions and Future Work}\label{sec:conclusion}

This paper presented \ac{NSuS}, a modular framework based on \ac{SuS}, that adapts niching ideas for use in rare-event simulation. When a performance function has challenging geometry, such as rapidly changing performance output or disconnected failure regions, \ac{NSuS} is able to produce more robust estimators for the probability of failure than \ac{SuS} is able to. This is due to its ability to avoid the ergodicity issues \ac{SuS} faces in these scenarios, which stem from the greedy nature of \ac{SuS}. Additionally, \ac{NSuS} can provide the user with at least some understanding of a reliability problem's topology. One of the main benefits of  \ac{SuS} when compared to other reliability methods is its ability to efficiently sample in high dimensions. This paper introduced the \ac{HVG}, a niching method specifically designed for high dimensions, so that \ac{NSuS} could retain this advantageous feature.

\ac{NSuS} is only able to detect and fully explore a niche if it initially produces some samples within that niche. This means that it is possible to construct counter-example performance functions for which \ac{NSuS} is not able to adequately populate the failure region. Of course, it is always possible to do this for any variance reduction technique, and across a range of practical scenarios, \ac{NSuS} has been shown to be more robust than \ac{SuS}. \ac{NSuS} could be made even more resilient by using it in conjunction with a Markov chain algorithm which encourages more exploration, so that all relevant niches can be initially populated. For the reliability problems with only one niche, where \ac{SuS} performs well, \ac{NSuS} will likely not branch and thus will perform identically to \ac{SuS}, aside from the additional computational cost incurred during the partitioner process. In these cases, \ac{SuS} is clearly the preferable algorithm, but this is impossible to know a priori, and for most situations the computational cost of the partitioner process is small compared to the cost of constructing the levels. The specific \ac{NSuS} implementation detailed in this paper does require the user to define some additional parameters, most notably the graph size. However, it was shown in the numerical examples that the performance of \ac{NSuS} is not particularly sensitive to the user chosen graph size.

The implementation of \ac{NSuS} described in this paper is only intended to serve as a straightforward baseline that can be widely applied. The process could be enhanced, or tailored specifically to incorporate knowledge about a specific reliability problem. For instance, the computational budget allocation could be changed so that some niches demand more resources than others, or that more resources are demanded overall. More complex hill valley tests could be used when constructing the \ac{HVG}. Alternative community detection algorithms and more sophisticated classification pipelines could be implemented. In particular, dimension reduction techniques could lower the computational cost of the \ac{HVG} construction and improve the performance of classification algorithms. An entirely different niching technique, adapted from existing \ac{EMO} literature, could also be considered. Another potential benefit of \ac{NSuS} that can be the subject of future work, is that the proposal samplers in different branches could be tuned independently to ensure maximum efficiency for each niche. The flexibility of \ac{NSuS} opens up a range of opportunities for future research and integration with other \ac{SuS} enhancements.

\newpage

\appendix

\section{Statistical Properties of Niching Subset Simulation}\label{appendix_a}

\let\originalthesection\thesection

\renewcommand{\thesection}{\Alph{section}}

\begin{proposition}\label{prop:unbias}
If an estimator $\hat{P}$ of $P$ is the sum of SuS estimators based on levels of size $n$, $\hat{P} = \sum_{t=1}^{T} \hat{P}_t$, where $\hat{P}_t$ is estimating $P_t$ and $P = \sum_{t=1}^{T} P_t $, then

\begin{equation}
    \left| \E\left[ \frac{\hat{P} - P}{ P}\right]\right| =  O(1/n).
\end{equation}
Thus $\hat{P}$ is asymptotically unbiased.
\end{proposition}

\begin{proof}
\begin{align*}
     \left| \E\left[ \frac{\hat{P} - P}{ P}\right]\right| &=  \left| \E\left[ \frac{\sum_{t=1}^{T} \hat{P}_t - P_t}{ \sum_{t=1}^{T} P_t}\right]\right| \\
     &\leq \left| \E\left[ \sum_{t=1}^{T} \frac{ \hat{P}_t - P_t}{  P_t}\right]\right| \\ 
     &= \left| \sum_{t=1}^{T} \E\left[  \frac{ \hat{P}_t - P_t}{  P_t}\right] \right| \\
     &\leq  \sum_{t=1}^{T} \left| \E\left[  \frac{ \hat{P}_t - P_t}{  P_t}\right] \right| \\
     &= O(1/n)
\end{align*}
where the last step is justified since the bias of a SuS estimator is $O(1/n)$ \cite{auEstimationSmallFailure2001}.
\end{proof}

\begin{proposition}\label{prop:consist}
If an estimator $\hat{P}$ of $P$ is the sum of SuS estimators based on levels of size $n$, $\hat{P} = \sum_{t=1}^{T} \hat{P}_t$, where $\hat{P}_t$ is estimating $P_t$, $P = \sum_{t=1}^{T} P_t $ and $\delta$ is the c.o.v. of $\hat{P}$, then 

\begin{equation}
    \delta^2 =  \E\left[ \frac{\hat{P} - P}{ P}\right]^2 = \sum_{i,j = 1}^T w_{ij} \delta_i \delta_j \rho_{ij} = O(1/n),
\end{equation}
where $w_{ij} = P_iP_j / \sum_{l,k=1}^T P_lP_k$, $\delta_t$ is the c.o.v. of $\hat{P}_t$ and $\rho_{ij}$ is the correlation between $\hat{P}_i$ and $\hat{P}_j$. It follows that $\hat{P}$ is a consistent estimator.
\end{proposition}

\begin{proof}
\begin{align*}
     \E\left[ \frac{\hat{P} - P}{ P}\right]^2 &=  \E\left[ \frac{( \sum_{t=1}^{T}\hat{P}_t - P_t)^2}{ (\sum_{t=1}^T P_t)^2}\right] \\
     &= \E \left[ \frac{\sum_{i,j=1}^{T} (\hat{P}_i - P_i)(\hat{P}_j - P_j)}{\sum_{i,j=1}^{T} P_i P_j}\right] \\ 
     &=  \E \left[ \sum_{i,j=1}^{T} w_{ij}\frac{ (\hat{P}_i - P_i)(\hat{P}_j - P_j)}{ P_i P_j}\right] \\ 
     &= \sum_{i,j = 1}^T w_{ij} \delta_i \delta_j \rho_{ij} \\
     &= O(1/n)
\end{align*}
where the last step is justified since the c.o.v. of a SuS estimator is $O(1/\sqrt{n})$ \cite{auEstimationSmallFailure2001}.
\end{proof}

\begin{proposition}\label{prop:cov_est}
If $\hat{P}_1 = \hat{P} \hat{P}_a $, $\hat{P}_2 = \hat{P} \hat{P}_b $ are estimators where $\hat{P}_a, \hat{P}_b$, and $\hat{P} $ are pairwise independent, $\delta, \delta_1$, and $\delta_2$ are the respective c.o.v. variables and $\rho$ is the correlation between $\hat{P}_1 $ and $\hat{P}_2 $, then

\begin{equation}
    \delta_1\delta_2 \rho = \delta^2.
\end{equation}
\end{proposition}

\begin{proof}
\begin{align*}
     \delta_1\delta_2 \rho &= \frac{\E[\hat{P}^2 \hat{P}_a \hat{P}_b ] - \E[\hat{P}\hat{P}_a ]\E[\hat{P}\hat{P}_b]}{\E[\hat{P}_1]\E[ \hat{P}_2]}\\
     &= \frac{\E[\hat{P}_a]\E[ \hat{P}_b] (\E[\hat{P}^2]-\E[\hat{P}]^2)}{\E[\hat{P}_a]\E[ \hat{P}_b] \E[\hat{P}]^2} \\ 
     &= \delta^2
\end{align*}
\end{proof}

\begin{proposition}\label{prop:sample}

Let $X \sim F_{X}$, $A_1, \ldots, A_n$ be a partition of the sample space, $X_i \sim X|A_i$, $\theta = (\theta_1, \ldots, \theta_n) \sim Multinomial(1;\prob(A_1), \ldots, \prob(A_n))$ and $Y = \sum_{i=1}^n \theta_i X_i \sim F_Y$. Then $X$ and $Y$ are identically distributed.

\end{proposition}

\begin{proof}
\begin{align*}
     F_{Y}(y) &= \prob(Y \leq y) \\ 
     &= \prob\left(\sum_{i=1}^{n} \theta_i X_i \leq y\right)  \\  
     &= \sum_{i=1}^{n} \prob(X_i \leq y) \prob(\theta_i = 1) \\
     &= \sum_{i=1}^{n} \prob(X \leq y | X \in A_i) \prob(A_i) \\
     &= \prob(X \leq y) \\
     &= F_{X}(y)
\end{align*}
\end{proof}

\let\thesection\originalthesection

\section{Hill Valley Graph Partitioner}\label{appendix_b}

\renewcommand{\thesection}{\Alph{section}}

\subsection{Asynchronous Label Propagation}\label{sec:alp}

Denote a graph as $G = (V,E)$, with vertices $V = \{x_1,\dots,x_n\}$. ALP updates the labels of the vertices iteratively through different discrete time steps. Let $C_{x_i}(t)$ denote the label of $x_i$ at time $t$. At the beginning of the algorithm each sample is assigned its own unique label, that is $C_{x_i}(0) = i$. To obtain the labels at step $t$ given the labels at step $t-1$, first the vertices are given a random ordering in which they shall be updated. Let $n_i$ be the number of vertices adjacent to $x_i$. In the specified order, node $x_i$ is updated with following rule:
\begin{equation}\label{eq:update rule}
    C_{x_i}(t) = h(C_{x_{i1}}(t),\dots,C_{x_{im}}(t),C_{x_{i(m+1)}}(t-1),\dots,C_{x_{in_i}}(t-1)),
\end{equation}
where $x_{i1},\dots, x_{im}$ are the vertices adjacent to $x_i$ that have already been updated, $x_{i(m+1)},\dots, x_{in_i}$ are the vertices adjacent to $x_i$ that are still awaiting an update for this iteration, and $h$ is function that returns the most common label. In the event of a tie, $h$ picks a label amongst those tied uniformly at random. Due to this tie breaking procedure it is inappropriate to have a stopping condition that looks for no label changes in an iteration, since it is always possible for some labels to change if there is a tie. Consequently, instead the algorithm stops if every vertex has a label that is amongst those labels that have caused a tie.

\subsection{Linear Support Vector Machine} \label{sec:lsvm}

Since it is possible for their to be more than two classes in the context of the \ac{HVG} partitioner, a one-vs-the-rest strategy is required, in which a classifier is created for each class. Formally, given a class, let $(\bm{x}_i,y_i)_{i=1}^n$ be the labelled samples, where $y_i = 1$ if $x_i$ is in the class and $y_i = -1$ otherwise. Now a confidence score is assigned to each sample, $\bm{w}^{T}\bm{x}_i + b$, where the coefficients $\bm{w} \in \R^d$ and intercept $b \in \R$ are determined by the following optimisation problem:
\begin{equation}\label{eq:slvm}
    \min_{\bm{w},b}\frac{1}{2}\rho(\bm{w}) + C \sum_{i=1}^n \max(0,1-y_i(\bm{w}^{T}\bm{x}_i + b))^{2},
\end{equation}
where $\rho$ is a penalty function and $C$ is regularisation parameter. Possible penalty functions include the L1 penalty, $\rho(\bm{w}) = \|\bm{w}\|_1$, and the L2 penalty, $\rho(\bm{w}) = \|\bm{w}\|_2^2$. Once this process has been completed for each class, each sample can be labelled according to whichever classifier gives it the largest confidence score. The examples in this paper use $C=1$ and $\rho(\bm{w}) = \|\bm{w}\|_1$.

\subsection{Balanced accuracy score} \label{sec:bas}

Let $(y_i,\hat{y}_i)_{i=1}^n$ be the true and predicted labels in a classification problem. Let the sample weights be defined as
\begin{equation}\label{eq:bas_weight}
    w_i = 1/\sum_{j=1}^{n}\one(y_j=y_i).
\end{equation}
The balanced accuracy score is then given as
\begin{equation}\label{eq:bas}
    \frac{1}{\sum w_i}\sum \one(\hat{y}_i=y_i) w_i.
\end{equation}

\newpage

\bibliography{references}

\end{document}